\documentclass[11pt,a4paper]{article}
\usepackage{jheparxiv}

\usepackage{amsmath}
\usepackage{amsfonts}
\usepackage{amssymb}
\usepackage{latexsym}
\usepackage{mathrsfs}
\usepackage{braket}		
\usepackage{slashed}
\usepackage{color}
\usepackage{xcolor}
\usepackage{slashed}
\usepackage{twistor}
\usepackage[all]{xy}
\usepackage{tikz-cd}
\usepackage{mathtools}
\usepackage{bm, bbm}
\usepackage{color}
\usepackage{braket}
\usepackage{float}
\usepackage[]{todonotes}
\usepackage{tikz-feynhand}
\usepackage{simplewick}
\usepackage{graphicx}
\makeatletter
\renewcommand{\todo}[2][]{%
    \@todo[caption={#2}, #1]{\begin{spacing}{0.5}#2\end{spacing}}%
} 
\makeatother 
\usepackage{setspace}

\title{All zeros of (super)String Theory}
\author[a]{Yu-Chi Chang,}
\emailAdd{r12222036@ntu.edu.tw}

\author[a]{Hsing-Yen Chen,} 
\emailAdd{B10202034@ntu.edu.tw}
\author[a,b]{and Yu-tin Huang}
\affiliation[a]{Department of Physics and Center for Theoretical Physics, National Taiwan University, Taipei
10617, Taiwan}
\affiliation[b]{Physics Division, National Center for Theoretical Sciences, Taipei 10617, Taiwan}
\emailAdd{yutin@phys.ntu.edu.tw}

\abstract{In this paper, we study the zeros of string theory utilizing its 
curve-integral representation. Firstly, we note that for bosonic strings the tachyon amplitude in curve-representation is identical to the kinematic shifted Tr$\phi^3$ amplitude. Scaffolding is then equivalent to taking the OPE limit of vertex operators on the string world-sheet, which yields amplitude of higher excitations. Using this picture, we derive that the $n$-point level-$N$ scattering amplitude shares the same set of zeros as the $n$-point tachyon amplitude as well as those inherited from it's prescaffold image, i.e. $(2^N)n$-point. Doing the same for the super-tachyon amplitude, exposes new zeros for the open super-string, which can be viewed as the avatar of supersymmetry. Finally we also consider the gluino amplitude at four and six-points, identifying its zero and recovering  super-Yang-Mills via scaffolding. Finally we consider the field theory limit of colored fermion amplitudes from the curve-integral form.     }

\begin{document}

\maketitle

\section{Introduction}
Recently, the geometric formulation of Tr$\phi^3$ amplitudes along with its stringy completion~\cite{Arkani-Hamed:2017mur, Arkani-Hamed:2019mrd, Arkani-Hamed:2019vag}, has remarkably found its foothold in a wide class of theories including Yang-Mills (YM) theory and none-linear sigma model (NLSM)~\cite{Arkani-Hamed:2023swr, Arkani-Hamed:2023jry, Arkani-Hamed:2024nhp}\,. 
The resulting ``curve-integral representation", where the amplitude is given as an integral over parameters associated with the triangulation of an $n$-gon surface, manifest many (hidden) properties of the amplitude including factorizations, hidden zeros, splitting near zeros and asymptotics~\cite{Arkani-Hamed:2023swr, Arkani-Hamed:2024fyd, Cao:2024gln, Cao:2024qpp, Arkani-Hamed:2024nzc}.

More precisely due to the $\text{SL(2,\,}\mathbb{R})$ invariance of the world-sheet integrand, the tree-level open-string amplitude can be naturally written in terms of cross-ratios: $u_{i,j}=\frac{z_{i{-}1,j}z_{i,j{-}1}}{z_{i,j}z_{i{-}1,j{-}1}}$. The positive parameterization of this space utilizes $n{-}3$ curve variables which can be associated with the internal cords of some triangulation of an $n$-gon. The triangulation itself, or more precisely the dual graph associated with the triangulation, provides the map between the $u$-variables and $y$-variables~\cite{Arkani-Hamed:2023lbd}. For example, the curve-integral representation of stringy Tr$\phi^3$ amplitude takes the following form:
\begin{equation}\label{eq: CurveInt}
\int \prod_{I=1}^{n{-}3}\frac{dy_I}{y_I} y^{\alpha'X_I} \prod_{i<j}(F_{i,j}(\mathbf{y_I}))^{-\alpha'c_{i,j}},
\end{equation}
where $y_I$ are the parameterization for the $n{-}3$ internal cords of the triangulation of $n$-gon, 
$F_{i,j}$ are polynomials of $y_I$s,  $c_{i,j}=-2p_i\cdot p_j$ and $X_{i,j}$s are defined through eq.(\ref{eq: CXMap}).  
The crucial step of connecting Tr$\phi^3$ to YM and NLSM is ``scaffolding''~\cite{Arkani-Hamed:2023swr} which involves the doubling of multiplicity such that the external data of the $2n$-amplitude is sufficient to parameterize the degrees of freedom for the $n$-point amplitude. Then taking pair-wise residues in $y$ reduces the $2n{-}3$ dimensional curve integral down to $n{-}3$. One of the important properties of curve-integral representation in eq.(\ref{eq: CurveInt}) is the existence of zeros. Depending on the explicit dependency of the polynomials $F_{i,j}$ on $y_I$s, taking a subset of $\alpha'c_{i,j}$ to be non-positive integers, $\alpha'c_{i,j}=-\mathbb{N}_0$, result in the curve integral to factorize a ``scaleless'' integral:
\begin{equation}
\int \frac{dy_I}{y_I}y^{\alpha' X_I{+}q},
\end{equation}
where $q$ is a non-negative integer. This type of integral vanishes after analytic continuation~\cite{Arkani-Hamed:2023swr}. These hidden zeros has profound implications, including the uniqueness of the field-theory limit scattering amplitude~\cite{Backus:2025hpn}, its connection to improved high-energy behavior~\cite{Rodina:2024yfc}. Their implications for massless non-ordered theories can be studied via double copy~\cite{Bartsch:2024amu, Li:2024qfp} (see \cite{Li:2024bwq, Zhang:2024efe, Huang:2025blb} for further studies.)

\begin{figure}[h!]
    \centering
    \includegraphics[width=0.45\linewidth]{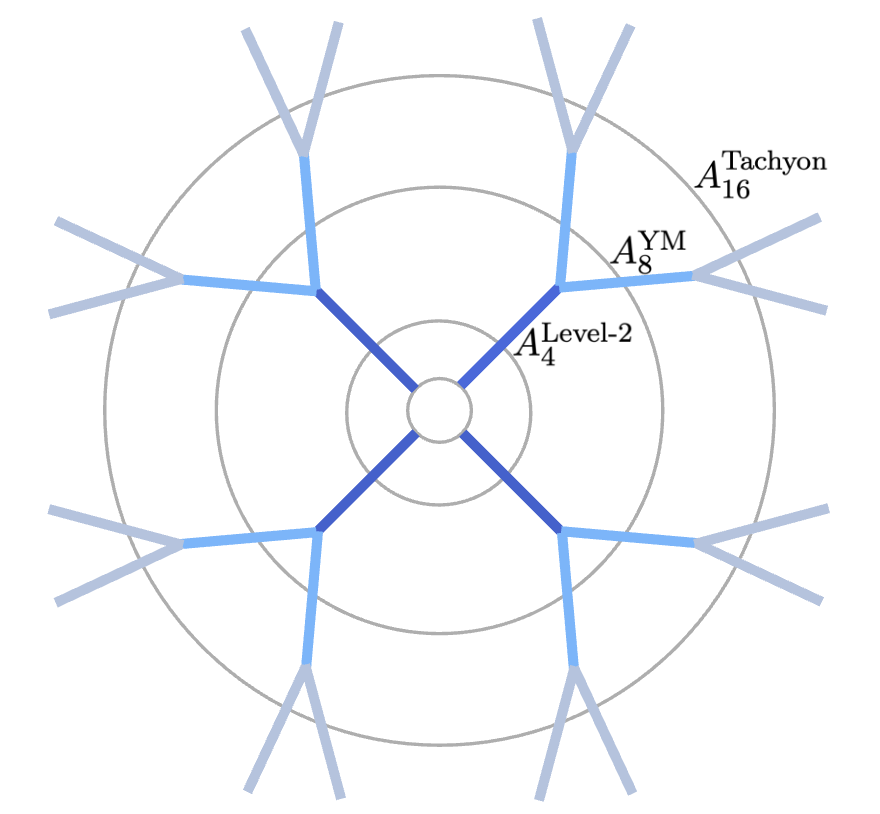} 
    \caption{Double scaffolding from 16-points tachyon amplitude to 4-points level-2.}
    \label{fig: 16to4Intro}
\end{figure}

In this paper we revisit open string amplitudes in the curve-integral representation. Firstly, we note that the tachyon amplitude takes the form:\begin{equation}
A^{\text{Tachyon}}_n= \int \prod_{I=1}^{n{-}3} \frac{dy_{I} }{y_{I}}\, y_{I}^{\alpha' X_{I}-1} \prod_{i<j}(F_{i,j}(\mathbf{y_I}))^{-\alpha'c_{i,j}},
\end{equation}
where $I$-denotes the internal cords that triangulate the surface. In this form, the $2n$-point integrand is identical to the kinematic shifted Tr$\phi^3$ amplitude. The only difference is that $X_{i,j}$ are now defined for massive momentum instead of massless. Thus, the shift that initiates the scaffolding is simply transforming the integrand into the tachyon amplitude. Indeed the next stage of scaffolding, which requires us to take the singularity of $X_{2i-1,2i{+}1}=0$, is ignorant to whether the original kinematics is massive or massless. Furthermore, as taking the residue of $X_{2i-1,2i{+}1}=0$ can be transmuted to taking the residue on the $y_I$ variables, the operation can be mapped to taking the residiue on the OPE of vertex operators. The fact that this yields the correct higher-level amplitudes is a simple consequence of the BRST invariant of the tachyon vertex operator. Namely, given a pair of conformal dimension 1 vertex operator $V$s, taking the residue of the OPE
\begin{equation}
\text{Res}_{z_1\rightarrow z_2}V(z_1)V(z_2)=V'(z_2)
\end{equation}
yields another conformal dimension 1 primary $V'$, i.e. the result must be BRST  and hence gauge invariant. Thus at tree-level, scaffolding of gluon \textit{is} massless factorization of the tachyon amplitude. This also suggests that one can iteratively scaffold amplitudes of  arbitrarily higher-level excitations. To simply put, for level-$N$ (with 1 being the massless gluons) take the $(2^N)n$-point tachyon amplitude and perform $N$-fold scaffolding. For example the 4-point massive spin-2 amplitude is then obtained via 2-fold scaffolding of the 16-point tachyon amplitude as illustrated in fig.\ref{fig: 16to4Intro}. Note that one \textit{could} have started at $2n$-point and directly scaffold arbitrary higher-level resonances, as these states must appear in the factorization channels at some points. However, as two particle kinematics are used to parameterize the degrees of freedom of the higher level resonance, i.e. components of their polarization tensor, it will be insufficient to disentangle the different irreps at a given mass level. 

The advantage of scaffolding is that it directly gives the higher-level amplitude in its curve-integral representation. For example from tachyon to gluon, one has:
\begin{align}
    A^{\text{Tachyon}}_{2n} = \int_0^\infty \prod_{s=1}^{n} \frac{dy_{2s-1,2s+1}}{y_{2s-1,2s+1}^2} y^{\alpha' X_{2s-1,2s+1}} \prod_{r=1}^{n-3} \frac{dy_{i_r,j_r}}{y_{i_r,j_r}^2} y^{\alpha' X_{i_r,j_r}} \prod_{i<j} F_{i,j}^{-\alpha' c_{i,j}}  \nonumber\\
    \xrightarrow{\quad\text{scaffold}\quad} \int_0^\infty \prod_{r=1}^{n-3} \frac{dy_{i_r,j_r}}{y_{i_r,j_r}^2} y^{\alpha' X_{i_r,j_r}}\left(\prod_{s=1}^n \partial_{y_{2s-1,2s+1}}\right)  \left( \prod_{i<j}F_{i,j}^{-\alpha' c_{i,j}} \right)\bigg|_{y_{2s-1,2s+1}=0},
\end{align}
where $a\in\{1,2,\dots,n\}$. While the resulting integrand might be complicated, especially if one scaffolds to higher levels, the pattern of zeros is transparent throughout the procedure. Firstly, if the amplitude before scaffolding is zero so will its descendents. Thus there are a set of zeros that are inherited from its $(2^N)n$ parent. Recall that the zeros of the curve integrals are determined by the polynomials $F_{i,j}$. Remarkably there is a simple relation between the polynomials of the parent $2n$-point surface and the  $n$-point:
\begin{equation}
\boxed{\quad F_{2i-1,2j-1}\bigl\lvert_{y_s=0}
=F_{2i,2j-1}\bigl\lvert_{y_s=0}
=F_{2i-1,2j}\bigl\lvert_{y_s=0}
=F_{2i,2j}\bigl\lvert_{y_s=0}
=F^{\rm amp}_{i,j}\quad},
\end{equation}
where the first four are $F$-polynomials of the parent $2n$-surface evaluated in the scaffolding limit $(y_s=0)$ and the last is the $F$-polynomial of an $n$-point surface, which we denote as $F^{\rm amp}$ indicating it corresponds to an amputation of the original $2n$-graph. This immediately allows us to conclude that the $n$-point scaffolded gluon amplitude enjoys the same zeros as the $n$-point tachyon amplitude. Iterating the argument to higher excitations, we see that the level-$N$ $n$-point scattering amplitude enjoys the same zeros as the $n$-point tachyon amplitude as well as those inherited from the $(2^N)n$-point parent amplitude. We demonstrate this is explicitly for the level-2 massive amplitude. 

All of the discussion for bosonic string is equally applicable for the superstring, where one can now choose to start with the NS-sector super-tachyon or the R-sector gluino. For the NS-sector, the $2n$-point super-tachyon there is a nice compact form utilizing grassmann-odd coordinates. It's conversion into curve integral is straight-forward and presented in eq.(\ref{eq: superTachyon}). See \cite{Cao:2025lzv} for recent work on relating these curve integrals to the Tr$\phi^3$ amplitude. We demonstrate that the super-tachyon enjoys the zeros of tachyon amplitude, with additional zeros stemming from the additional pfaffian factors. For example, at $2n$-point the amplitude will vanish if we set \(c_{i,j} = 0\)  for both even (or odd) indices.
These zeros can be viewed as a hallmark of supersymmetry. Not surprisingly upon scaffolding, these zeros lead to the well known zeros of supersymmetric YM theory, i.e. the amplitude vanishes when we set all inner product of polarization vextors to zero ($\epsilon_i\cdot \epsilon_j=0$) the amplitude vanishes. For the R-sector, the ground state is gluinos. Using previously known results, we present the four- and six-point amplitudes in curve-integral representation. We deduce that the four- and six-point gluino amplitudes share the same zeros as the tachyon amplitude.

\section{Lightning review of curve-integral representation Tr$\phi^3$.}
It has proven extremely enlightening to encode the kinematics of massless planar scattering amplitudes, where there is a natural ordering of the external legs, using the ``kinematic mesh"~\cite{Arkani-Hamed:2019vag}. While the setup might be familiar to many by now, for completeness we present a self-contained review.

\begin{figure}[h!]
    \centering
    \begin{minipage}{0.3\linewidth}
        \centering
        \includegraphics[width=\linewidth]{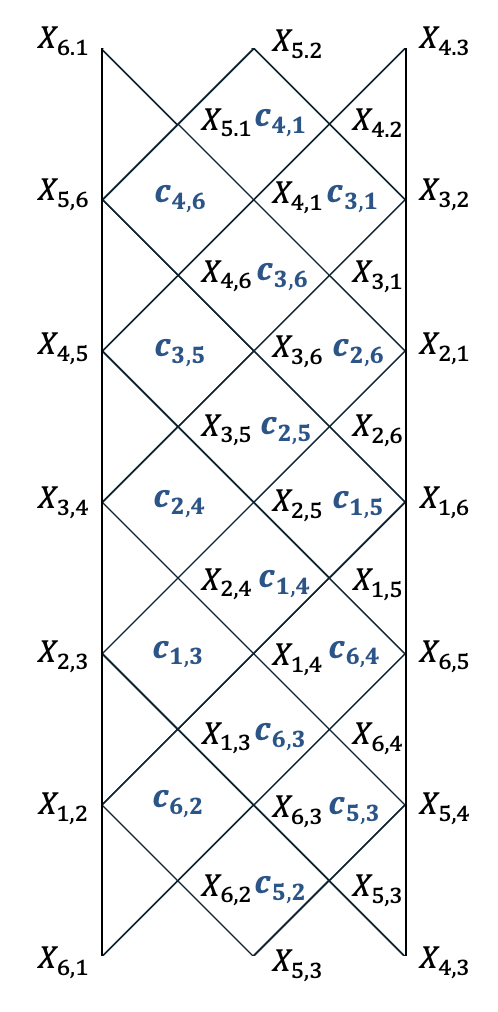}
        \caption{The mesh diagram at $6$-points that conveniently captures the relation in eq.(\ref{eq: CXMap}). }
        \label{111}
    \end{minipage}%
    \hspace{0.05\linewidth}
    \begin{minipage}{0.3\linewidth}
        \centering
        \includegraphics[width=\linewidth]{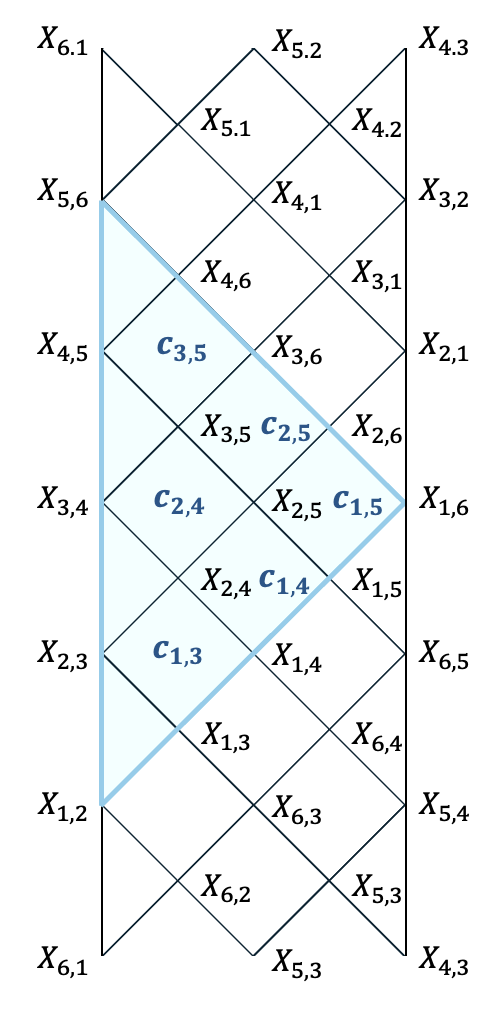}
        \caption{The independent kinematics for the ray-like triangulations at $6$-points.}
        \label{123}
    \end{minipage}
\end{figure}

Starting with $X_{i,j}\equiv (p_i{+}p_{i{+}1}{+}\cdots p_{j{-}1})^2$. From its definition, we have $X_{i,j}=X_{j,i}$. All kinematic invariants can be expressed in terms of these ``dual invariants" via
\begin{equation}\label{eq: CXMap}
c_{i,j}\equiv -2 p_i\cdot p_j=X_{i,j}{+}X_{i{+}1,j{+}1}-X_{i,j{+}1}-X_{i{+}1,j}\,.
\end{equation}
Importantly, this relation holds for general kinematics, irrespective of $p_i^2$. Thus any scalar amplitude, with or without ordering, can be expressed in terms of $X_{i,j}$s. The mesh diagram is given in fig.\ref{111}, where we start with two vertical lines marked with $X_{i,i{+}1}$, and we connect the points with 45$^o$ lines. The remaining $X_{i,j}$s resides on the vertices of, where the right (left) labels increase by 1 as one ascends along the upper-left (right) direction. Each square tile then represents the sum of the upper lower vertex minus the left right vertex, i.e. $c_{i,j}$ in eq.(\ref{eq: CXMap}). 

Let's begin with the stringy completion of Tr$\phi^3$ theory:
\begin{equation}
A^{\rm Tr\phi^3}_n=\int_{z_1<\cdots<z_n}\frac{dz_1\cdots dz_n}{\text{SL(2,\,}\mathbb{R})\mathbb{PT}}\prod_{i<j}z^{-\alpha'c_{i,j}}_{i,j}\,,
\end{equation}
where $\mathbb{PT}$ is the ``Park-Taylor factor" $\prod_{i}z_{i,i{+}1}$ and \(z_{i,j}:=z_j - z_i,\). The $z$ variables can be understood as a positive parameterization of the $u$-variables. Starting with an $n$-gon. For each of the $n(n{-}3)/2$ internal edges assign a variable $u_{i,j}$. The $u$-variables satisfies the $u$-equations: 
\begin{equation}\label{eq: ueq}
u_{i,j}+\prod_{(k,l)\cap (i,j)}u_{k,l}=1\,.
\end{equation} The $n{-}3$-dimensional space defined via $0<u_{i,j}<1$ subject to eq.(\ref{eq: ueq}) is a positive geometry denoted as $U_n^+$. The relation between the $u$- and the $z$- variables is via the cross-ratio map:
\begin{equation}\label{eq: uz}
u_{i,j}=\frac{z_{i{-}1,j}z_{i,j{-}1}}{z_{i,j}z_{i{-}1,j{-}1}}\,,
\end{equation}
where the fact that $z_i$s are real and ordered ensures that $0<u<1$. In the $u$-variables the Koba-Nileson factor becomes $\prod_{i<j}z^{-\alpha'c_{i,j}}_{i,j}=\prod_{i<j}u_{i,j}^{\alpha'X_{i,j}}$. The remaining measure $\int\frac{dz_1\cdots dz_n}{\text{SL(2,\,}\mathbb{R})\mathbb{PT}}$ is the canonical form for $U_n^+$. Note that the dimension of $U_n^+$ exactly matches the number of propagators in an $n$-point cubic diagram. This suggests a parametrization that can be graphically represented by an $n$-point Feynman diagram, a Feynman ``fatgraph". In the following we will closely adhere to the construction in \cite{Arkani-Hamed:2024nzc} which is slightly modified from~\cite{Arkani-Hamed:2023lbd}.

Begin with an $n$-gon with its corner labeled successively such that each edge is given as $(i,i{+}1)$. A given triangulation, denoted as $\mathcal{T}$, is defined by the $n{-}3$ cords that dissect the internal area. The dual fatgraph has $n{-}3$ ``propagators" and we label the $i$-th external leg as the one that flows out of edge $(i,i{+}1)$. Now for the propagator that is intersected by the cord $(i,j)$ we associate a variable $y_{i,j}$. For a given fatgraph there are ``curves" $(i,j)$ that are paths that connects legs $i$ and $j$. For tree fatgraphs, which is all that is relevant for this paper, the path is unique, and we can associate a ``word" built out of the binary alphabet $L/R$ which records the successive left/right turn taken along the path. We will trim this word by removing any continuous right turns from the beginning and continuous left turns from the end.
From the final word we can construct the $F$-polynomial as follows: starting from external leg $i$ (or entering from edge $(i,i{+}1)$) the $L/R$ choices is translated into two $2\times2$ matrix:
\begin{equation}
M_{L}(y)=\begin{pmatrix}
y & y \\
0 & 1 
\end{pmatrix},\quad M_{R}(y)=\begin{pmatrix}
y & 0 \\
1 & 1 
\end{pmatrix}\,,
\end{equation}
where $y$ is the variable associated with each propagator. Thus a curve $(i,j)$ in a given triangulation consists of $(n{-}3)$ left/right turns, from which we can build the polynomial $F_{i{-}1,j{-}1}$:
\begin{equation}
F_{i{-}1,j{-}1}=(1,1)\cdot M_{L/R}(y_1)\cdots M_{L/R}(y_{n{-}3})\cdot \begin{pmatrix}
1  \\
0  
\end{pmatrix}
\end{equation}
The quantity $F_{i,j}$ is equal to one whenever $(i+1,j+1)$ is one of the cords in $\mathcal{T}$ or when the indices satisfy $|i-j| \leq 1$.

Now we are ready to give the parameterization of $u$-variables in terms of $y_I$ variables. Given a $u_{i,j}$ we consider the corresponding curve $(i,j)$ considering two distinct cases:

\paragraph{Case 1: $(i,j) \in \mathcal{T}$}
For the set of indices belonging to a cord of the chosen triangulation $\mathcal{T}$, we have:
\begin{equation}
\label{eq:u_from_F_T}u_{i,j} = y_{i,j}\,\frac{F_{i-1,j}\,F_{i,j-1}}{F_{i,j}\,F_{i-1,j-1}}.
\end{equation}

\paragraph{Case 2: $(i,j) \notin \mathcal{T}$}
For indices outside the triangulation $\mathcal{T}$, the relationship simplifies to:
\begin{equation}
\label{eq:u_from_z}u_{i,j} = \frac{F_{i-1,j}\,F_{i,j-1}}{F_{i-1,j-1}\,F_{i,j}}.
\end{equation}

Using these transformations, it was shown in~\cite{Arkani-Hamed:2023swr} that the stringy Tr$\phi^3$ amplitude takes the form: 
 \begin{eqnarray}\label{eq: Trphi3y}
A^{\rm Tr\phi^3}=\int_{z_1<\cdots<z_n}\frac{dz_1\cdots dz_n}{\text{SL(2,\,}\mathbb{R})\mathbb{PT}}\prod_{i<j} u_{i,j}^{\alpha' X_{i,j}}=\int_{\mathbb{R}_{>0}}\prod_{I=1}^{n{-}3} \frac{dy_I}{y_I} y_I^{\alpha'X_I}\prod_{i<j}(F_{i,j}(\mathbf{y_I}))^{-\alpha'c_{i,j}}\,,
\end{eqnarray}
where $y_I$ are the propagator variables associated with the fatgraph of a given triangulation. Note that when written in terms of $u$-variables alone, it makes obvious that the singularities of the amplitude can only be of planar poles, i.e. $X_{i,j}$s. Indeed the $u$-variables were widely used in the early days of dual resonance models~\cite{Koba:1969rw, Bardakci:1968rse, Chan:1969ex, Gross:1969db}. Since $F_{i,i{+}1}=1$, the kinematics dependence of the integrand is given in terms of the $n{-}3$ $X_{i,j}$ associated with the internal cords, which we will denote as $X_{I}$ and the remaining $n(n{-}3)/2-(n{-}3)=(n{-}3)(n{-}2)/2$ invariants are precisely the non-planar variables $c_{i,j}$ where $(i{+}1,j{+}1)$ are not one of the cords and $|i{-}j|>$1.

Consider a ray like triangulation, with the internal cords given by $X_{1,i}$ where $2<i<n$. The variables $y_{1,i}$ then only appear in $F_{a,b}$ with $1\leq a\leq i{-}2$ and $i\leq b\leq n{-}1$. Thus by setting the exponents of these $F_{a,b}$s to non-negative integer, i.e. $\alpha'c_{a,b}=-\mathbb{N}_0$, where $\mathbb{N}_0$ is non-negative integer, the integral for $y_{1,i}$ reduces to: 
\begin{equation}
\int_{\mathbb{R}_{>0}} \frac{dy_{1,i}}{y_{1,i}}y_{1,i}^{\alpha'X_{1,i}{+}q},
\end{equation}
where $q$ is a non-negative integer. As discussed in~\cite{Arkani-Hamed:2023swr}, these are scaleless integrals and hence integrates to zero.

\section{Curve-integral representation for Bosonic string and scaffolding}
We start by considering the $2n$-point tachyon scattering amplitude in the bosonic string:
\begin{equation}
A^{\text{Tachyon}}_{2n} 
= \int \frac{dz_1\,dz_2\cdots dz_{2n}}{\mathrm{SL}(2,\mathbb{R})}\,\Bigl\langle\prod_{a=1}^{2n} e^{i p_a\cdot X(z_a)}\Bigr\rangle =\int \frac{dz_1\cdots dz_{2n}}{\text{SL(2,\,}\mathbb{R})}\prod_{i<j} z_{i,j}^{-\alpha' c_{i,j}},
\end{equation}
where we now have \(p_i^2 = \tfrac{1}{\alpha'}\). We now convert the integrand to $u$-variables. Firstly, due to massive kinematics we have the following identities:
\begin{equation}\label{eq: CurveId}
\prod_{C} u_{C}=\prod_i\frac{z_{i,i{+}1}}{z_{i,i{+}2}},\quad\quad  \prod_{C} u_{C}^{\alpha' X_{C}} = \prod_i^n z_{i,i+1}^{2\alpha'p_i^2}\prod_i^n z_{i,i+2}^{-\alpha'p_i^2}\prod_{i<j} z_{i,j}^{-\alpha' c_{i,j}}\,,
\end{equation}
where $C$ represent all internal cord of an $n$-gon, i.e. $i,j$ with $i<j-1$. Then the $n$-tachyon amplitude becomes~\cite{Arkani-Hamed:2023jry}:
\begin{align}\label{eq: TachyonY}
A^{\text{Tachyon}}_n = \int \frac{dz_1\cdots dz_{n}}{\mathrm{SL}(2,\mathbb{R})\mathbb{PT}}\,\mathbb{PT}\prod_{i<j} z_{ij}^{-\alpha' c_{i,j}}= \int_0^\infty \prod_{I} \frac{dy_{I} }{y_{I}} \prod_C u_C^{\alpha' X_C - 1}\,,
\end{align}
where we've used the fact that $\int\frac{dz_1\cdots dz_n}{\text{SL(2,\,}\mathbb{R})\mathbb{PT}}$ is a canonical form of $U_n^+$ and replaced it with the positive parameterization associated with some chosen triangulation. Using eq.(\ref{eq:u_from_F_T}) and (\ref{eq:u_from_z}), we find:
\begin{equation}
A^{\text{Tachyon}}_n= \int \prod_{I} \frac{dy_{I} }{y_{I}}\, y_{I}^{\alpha' X_{I}-1} \prod_{i<j}(F_{i,j}(\mathbf{y_I}))^{-\alpha'c_{i,j}}\,.
\end{equation}
Compared to eq.(\ref{eq: Trphi3y}), we see that aside from the extra $y^{{-}1}_{I}$ the two integrands are identical. An immediate consequence is that the zeros of Tr$\phi^3$ are identical to the Tachyon amplitude. Indeed these zeros were found long ago~\cite{DAdda:1971wcy}. 
Furthermore for even multiplicity $2n$, the integrand is exactly the ``deformed" Tr$\phi^3$ integrand. In particular starting with eq.(\ref{eq: Trphi3y}), by performing the kinematic shift $\alpha'X_{e,e}\rightarrow \alpha'X_{e,e}+1$ and $\alpha'X_{o,o}\rightarrow \alpha'X_{o,o}+1$, where $e,e$ and $o,o$ represent double even and double odd indices, respectively. The shifted amplitude takes the form:
\begin{equation}
A^{\rm Tr\phi^3}_{2n,{\rm deform}}=\int_{z_1<\cdots<z_n}\frac{dz_1\cdots dz_n}{\text{SL(2,\,}\mathbb{R})\mathbb{PT}}\prod_{C} u_{C}^{\alpha' X_{C}}\frac{\prod u_{e,e}}{\prod u_{o,o}}=\int \prod_{I} \frac{dy_{I} }{y^2_{I}}\, y_{I}^{\alpha' X_{I}} \prod_{i<j}(F_{i,j}(\mathbf{y_I}))^{-\alpha'c_{i,j}},
\end{equation}
which is the same as eq.(\ref{eq: TachyonY}). Thus we see that $A^{\rm Tr\phi^3}_{2n,{\rm deform}}$ and $A^{\rm Tachyon}_{2n}$ takes the same form, once the kinematics is cast into a particular triangulation basis.
Note that this is of course only true in this representation, where $X_{i,j}$ corresponds to massless kinematics for Tr$\phi^3$ and massive for tachyons.\footnote{More precisely, while eq.(\ref{eq: CXMap}) holds for both cases, $X_{i,i{+}1}=0$ for massless kinematics and $X_{i,i{+}1}=\frac{1}{\alpha'}$ for tachyons.}

In general, converting the world-sheet integrals into its' curve-integral representation can be an arduous task, as can be seen in our later derivation of the six-point gluino amplitude. Essentially, one needs to rewrite the world-sheet integrand into $\text{SL(2,\,}\mathbb{R})$ invariant blocks, before converting to $u$-variables via eq.(\ref{eq: uz}). While for massless amplitudes, gauge invariance provides a natural guiding principle for organizing $\text{SL(2,\,}\mathbb{R})$ invariant blocks, beyond massless level a systematic organization is still lacking.  On the otherhand starting with the $2n$-point tachyon amplitude, taking successive two-particle factorization limits one recovers amplitudes of higher-level excitations in the curve-integral representation.

 Note that since $A^{\rm Tachyon}_{2n}$ and $A^{\rm Tr\phi^3}_{2n,{\rm deform}}$ has the same integral representation, the scaffolding procedure for Yang-Mills amplitude can be simply understood as factorization limits of the Tachyon amplitude.

\subsection{From Tachyon to Yang-Mills amplitudes}

Let us review the scaffolding procedure by extracting the $n$-point Yang–Mills amplitude from the $2n$-point Tachyon amplitude. Denoting the $2n$-point momenta as $\{p_i\}$ and $n$-point as momenta $\{k_i\}$, we identify:
\[
p_{2i-1} + p_{2i} = k_i, \qquad i=1,2,\dots,n.
\]
We extract the residue of $A^{\mathrm{Tachyon}}_{2n}$ on the massless factorization pole defined by:
\begin{equation}
X_{2i{-}1,2i} := (p_{2i{-}1} + p_{2i})^2 = k_i^2=0 \quad\Longrightarrow\quad c_{2i{-}1,2i} = \tfrac{2}{\alpha'},
\end{equation}
since $p_{2i-1}^2 = p_{2i}^2 = \tfrac{1}{\alpha'}$. 

Suppose we want to extract the residue of the amplitude at the pole $\alpha'c_{2i{-}1,2i}=n$, as the only non-analytic dependence on $c_{2i{-}1,2i}$ stems from the exponents of \( z_{2i-1,2i} \), the singularity must be correlated with the limit \( z_{2i-1,2i}\rightarrow 0 \). Focusing on the $z_{2i{-}1,2i}$ factor we write the amplitude as:
\begin{equation}
\int dz_{2i{-}1}\,dz_{2i}\, z_{2i{-}1,2i}^{-\alpha' c_{2i{-}1,2i}} F(z_{2i-1}, z_{2i}, \dots).
\end{equation}
Following~\cite{Arkani-Hamed:2023jry}, we perform the following change of variables:
\[ z_{2i-1,2i}=U, \quad z_{2i-1}=V, \quad z_{2i}=U+V, \quad dz_{2i-1}\,dz_{2i}=dU\,dV. \]
Then the integrand becomes:
\[
\int dU\,dV\, U^{-\alpha' c_{2i{-}1,2i}}\, F( V, U+V, \dots).
\]
Since the relevant part of the integral that contributes to any potential singularity in $c_{2i{-}1,2i}$ is around $U=0$, we find:
\begin{equation}
\begin{split}
    &\int dV\int_0^\delta  dU\, U^{ -\alpha' c_{2i{-}1,2i}}  
    \left( 
        F(V, V) 
        + U\,\frac{\partial F}{\partial U} \Big|_{U=0} 
        + \frac{U^2}{2!}\,\frac{\partial^2 F}{\partial U^2} \Big|_{U=0}
        + \cdots 
    \right) \\[6pt]
    &=\int dV \Bigg[
        \frac{F|_{U=0}}{-\alpha'c_{2i{-}1,2i}+1}\, 
        + \frac{\frac{\partial F}{\partial U} |_{U=0}}{-\alpha'c_{2i{-}1,2i}+2}\,
        + \cdots 
        + \frac{1}{(n-1)!}\, \frac{\frac{\partial^{n-1} F}{\partial U^{n-1}} \Big|_{U=0}}{-\alpha'c_{2i{-}1,2i}+n}
    \Bigg]. \\[6pt]
   \end{split}
\end{equation}
Thus, we conclude that taking the residue at the pole \( \alpha'c_{2i{-}1,2i}= n \) is equivalent to simply setting $\alpha'c_{2i{-}1,2i}=n$ on the integrand and taking \( z_{2i-1,2i} = 0 \) residue.

We now demonstrate that one reproduces the Yang-Mills correlator on the factorization of the massless pole, i.e. $\alpha' c_{2i{-}1,2i}\rightarrow 2$:
\begin{equation}
\mathrm{Res}_{\alpha' c_{2i{-}1,2i}\rightarrow 2}\,A^{\mathrm{Tachyon}}_{2n}
= \mathrm{Res}_{\alpha' c_{2i{-}1,2i}\rightarrow 2}
\int \frac{dz_1\cdots dz_{2n}}{\text{SL(2,\,}\mathbb{R})}\,
 z_{2i-1,2i}^{-\alpha' c_{2i{-}1,2i}} \prod_{(a,b)\neq(2i-1,2i)} z_{a,b}^{-\alpha' c_{a,b}}.
\end{equation}
Denoting $\prod_{(a,b)\neq(2i-1,2i)} z_{a,b}^{-\alpha' c_{a,b}}$ as $F$, we obtain:

\begin{equation}
\begin{aligned}
\mathrm{Res}_{\alpha' p_{2i{-}1}p_{2i}\rightarrow {-}1}\,A^{\mathrm{Tachyon}}_{2n}= \int \frac{dz_1\cdots  \,dV\cdots dz_{2n}}{\text{SL(2,\,}\mathbb{R})}
\,\partial_{U}F\big|_{U=0}\,.
\end{aligned}
\end{equation}
Note that since 
\begin{equation}
\partial_{U}F\big|_{U=0}=\bigl\langle \,p_{2i}\!\cdot i\partial X(z_{2i})\,e^{i(p_{2i-1}+p_{2i})\!\cdot\!X(z_{2i})}\prod_{a\neq2i}e^{i p_a\cdot X(z_a)}\bigr\rangle,
\end{equation}
we find that by setting $p_{2i}=\epsilon_i$, and $p_{2i-1}+p_{2i}=k_i$, we recover:
\begin{equation}
\int \frac{dz_1\cdots dz_{2i-2}\,dz_{2i}\cdots dz_{2n}}{\text{SL(2,\,}\mathbb{R})}
\bigl\langle \,\epsilon_{i}\cdot i\partial X(z_{2i})\,e^{ik_i\!\cdot\!X(z_{2i})}\prod_{a\neq2i}e^{i p_a\cdot X(z_a)}\bigr\rangle\,,
\end{equation}
i.e. the correlator of a Yang-Mills vertex operator with $2n-1$ tachyons. Note that the physical condition $k_i\cdot\epsilon_i=0$ is automatically satisfied. By imposing the condition $\alpha'c_{2i{-}1,2i}=2$ for all $i=1,\dots,n$, we obtain the $n$-point Yang-Mills amplitude:
\[
\int \frac{\prod_{a=1}^n dz_{2a}}{\text{SL(2,\,}\mathbb{R})}
\Bigl\langle \prod_{i=1}^n\bigl(\epsilon_{i}\cdot i\partial X(z_{i})\bigr)
\,e^{i k_i\cdot X(z_{i})}\Bigr\rangle,
\]
thus completing the scaffolding procedure from tachyon to Yang-Mills amplitude.

To reach the next level, there are two choices. One can start with $2n$ tachyons and take $c_{2i{-}1,2i}=3$ residue, or starting with $4n$ tachyon amplitude, scaffold to Yang-Mills and then to level 2. Since we would like to have sufficient independent momenta in order to span the polarization tensors of level $2$, we will proceed with later approach.   
\subsection{From Yang-Mills to Level-2 amplitudes}
\label{scaffoldcorrelator}
We can continue the scaffolding procedure to higher levels. Let's consider the construction of $n$-point spin-2 (level-2) amplitude from $2n$-point Yang-Mills amplitude:
\begin{equation}\label{eq: YM}
    A^{\,\text{YM}}_{2n}=\int \frac{\prod_{a=1}^{2n} dz_a}{\text{SL(2,\,}\mathbb{R})}
    \left\langle \prod_{a=1}^{2n}\epsilon_a\cdot i \partial X(z_a)
    \,e^{i p_a\cdot X(z_a)}\right\rangle.
\end{equation}
Since we are isolating the massive level-2 resonance with $m^2 = \tfrac{1}{\alpha'}$ from massless external legs, the kinematic limited is taken as, again denoting $k_i=p_{2i{-}1}{+}p_{2i}$:
\[
    k_i^2 = (p_{2i-1}{+}p_{2i})^2 \quad\Longrightarrow\quad -\frac{1}{\alpha'}.
\]
As we will be taking the residue at $z_{2i{-}1,2i}\rightarrow 0$, we first examine the different contractions in eq.(\ref{eq: YM}) that leads to such pole. Focusing on the singularity associated with $z_{1,2}$, let us organize the correlator into different powers of $z_{1,2}$ associated with different ways of contracting $\partial X$, as these factors can only contract once. There are five types of dependence:
\begin{equation}
\left( \frac{(\epsilon_1 \cdot p_2)(\epsilon_2 \cdot p_1)}{z_{1,2}^2},\quad  \frac{\epsilon_2 \cdot \epsilon_1}{z^2_{1,2}},\quad   \frac{\epsilon_2 \cdot p_1}{z_{1,2}},\quad  \frac{\epsilon_1 \cdot p_2}{z_{1,2}},\quad {\rm others}  \right),
\end{equation}
where the first and second term corresponds to the  $\partial X$ factor of vertex (2)1 contracted with vertex (1)2, the third and fourth term has one  $\partial X$ contracted with other vertex operators and the last term has both $\partial X$ contracted with others. The overall KN factor contains $z^{-\alpha'c_{1,2}}_{1,2}$, which is evaluated in $c_{1,2}\rightarrow \frac{1}{\alpha'}$ gives an additional $z^{{-}1}_{1,2}$. Expanding in the limit $z_2\rightarrow z_1$ and taking the residue we obtain:

\begin{equation}
\left\langle 
\left[
\begin{aligned}
&\left( - 2 \alpha'^2 (\epsilon_1 \cdot p_2)(\epsilon_2 \cdot p_1)+\alpha' (\epsilon_1 \cdot \epsilon_2)\right)\\
&\qquad\times(p_2 \cdot i\partial^2 X+(p_2 \cdot  i\partial X)(p_2 \cdot i\partial X) )  \\
&- 2 \alpha' (\epsilon_1 \cdot p_2) (\epsilon_2 \cdot i \partial^2 X + (\epsilon_2 \cdot i\partial X)(p_2 \cdot i\partial X)) \\
&+ 2 \alpha' (\epsilon_2 \cdot p_1) (\epsilon_1 \cdot i\partial X)(p_2 \cdot i \partial X) \\
&+ (\epsilon_1 \cdot i \partial X)(\epsilon_2 \cdot i \partial X)
\end{aligned}
\right]
e^{i k_1 \cdot X(z_2)} 
\prod_{i=3}^{2n} \epsilon_i \cdot i \partial X(z_i) e^{i p_i \cdot X(z_i)} 
\right\rangle.   
\end{equation}
We can readily identify the level-2 vertex operator to take the form:
\begin{equation}
V(z)_{N=2} = \left[ B^\mu \, i \partial^2 X_\mu(z) + E^{\mu\nu} \, i \partial X_\mu(z) \, i \partial X_\nu(z) \right] e^{i k \cdot X(z)},
\end{equation}
with
\begin{equation}
\begin{aligned}
&B^\mu = \left( -2\alpha'^2 (\epsilon_1 \cdot p_2)(\epsilon_2 \cdot p_1) + \alpha' (\epsilon_1 \cdot \epsilon_2) \right) p_2^\mu 
- 2 \alpha' (\epsilon_1 \cdot p_2) \, \epsilon_2^\mu, \\
&E^{\mu\nu} = 
\bigg[
\left( -2 \alpha'^2 (\epsilon_1 \cdot p_2)(\epsilon_2\cdot p_1) + \alpha' (\epsilon_1 \cdot \epsilon_2) \right) (p_2^\mu\otimes p_2^\nu)  -  \alpha' (\epsilon_1 \cdot p_2) \, \left( (\epsilon_2^\mu \otimes p_2^\nu)+(p_2^\mu \otimes\epsilon_2^\nu)\right)\\
&\hspace{12em}+ \alpha' (\epsilon_2 \cdot p_1) \left( (\epsilon_1^\mu\otimes p_2^\nu) + (p_2^\mu \otimes \epsilon_1^\nu) \right) + \frac{1}{2}\left( (\epsilon_1^\mu \otimes \epsilon_2^\nu) + (\epsilon_2^\mu \otimes \epsilon_1^\nu )\right)
\bigg],
\label{polBE}
\end{aligned}
\end{equation}
where we have symmetrized $E^{\mu\nu}$.

We can check whether our newly identified $E^{\mu\nu}$ and $B^\mu$ satisfies the requisite constraint imposed by BRST invariance. Recall that BRST invariance requires the vertex operator to be a primary of weight 1. Taking the OPE of $V_{N=2}$ with the stress tensor $T_X=-\frac{1}{2}\eta_{\rho\sigma}\partial X^\rho \partial X^\sigma$, we find:
\newcommand{\nor}[1]{:\mathrel{#1}:}
\begin{equation}
\begin{split}
    T_X(z) V_{N=2}(0) &= {\frac{1}{2}\eta_{\rho\sigma}i\partial X^\rho i\partial X^\sigma(z)} {\Big(B^\mu i\partial^2X_\mu e^{ik \cdot X}+E^{\mu\nu} i\partial X_\mu i\partial X_\nu e^{ik\cdot X}\Big)(0)} \\
    &= \left(8\alpha'^2(k\cdot B)+4\alpha'^2\text{Tr}(E)\right) \frac{ e^{ik\cdot X}}{z^4} + \left(4\alpha' (B \cdot i\partial X)+8\alpha'^2 E_{\mu\nu} k^\mu i\partial X^\nu \right)\frac{e^{ik\cdot X}}{z^3} \\
    &\qquad +\mathcal{O}(\frac{1}{z^2})+ \mathcal{O}(\frac{1}{z}) + \text{regular}.
\end{split}
\end{equation}
To ensure that the vertex operator $V_{N=2}$ is a primary field, we can derive the corresponding constraints on the polarization vector $B^\mu$ and tensor $E^{\mu\nu}$:
\begin{equation}
\begin{split}
    &8\alpha'^2(k\cdot B)+4\alpha'^2\text{Tr}(E)=0,\\
    &4\alpha' B^\nu+8\alpha'^2  k_\mu E^{\mu\nu} =0.
\label{polconstraint}
\end{split}
\end{equation}
We then substitute the polarization vector and tensor obtained from scaffolding eq.(\ref{polBE}) into the constraints eq.(\ref{polconstraint}) and find that it is indeed satisfied, where we have used $k^\mu=p^\mu_1+p^\mu_2$ and $c_{1,2}=\tfrac{1}{\alpha'}$. From this, we conclude that the polarization vector and tensor extracted from the scaffolding vertex operator automatically satisfy the BRST constraint. Note that for convenience, we can choose the gauge $\epsilon_1\cdot p_2=\epsilon_2\cdot p_1=0$, then we simply have:
\begin{equation}
B^\mu = \alpha' (\epsilon_1 \cdot \epsilon_2)  p_2^\mu,\quad E^{\mu\nu} = 
 \alpha' (\epsilon_1 \cdot \epsilon_2)  (p_2^\mu\otimes p_2^\nu)+ \frac{1}{2}\left( (\epsilon_1^\mu \otimes \epsilon_2^\nu) + (\epsilon_2^\mu \otimes \epsilon_1^\nu )\right)\,.
\end{equation}

Finally, we take the residues at $z_{2i-1,2i} \rightarrow 0$ for $i = 1, \ldots, n$ to obtain the $n$-point level-2 amplitude from the Yang–Mills amplitude and complete the scaffolding process:
\begin{equation}
\begin{aligned}
&\mathrm{Res}_{\alpha' c_{2i{-}1,2i}\rightarrow 1}\,A^{\mathrm{YM}}_{2n}= \mathrm{Res}_{\alpha' c_{2i{-}1,2i}\rightarrow 1}\,\int \frac{\prod_{a=1}^{2n} dz_a}{\mathrm{SL}(2,\mathbb{R})}
    \left\langle \prod_{a=1}^{2n}\epsilon_a\cdot i \partial X(z_a)
    \,e^{i p_a\cdot X(z_a)}\right\rangle \\
    &=\int \frac{\prod_{a=1}^n dz_{2a}}{\mathrm{SL}(2,\mathbb{R})}
\Bigl\langle \prod_{i=1}^n\left[ B_i^\mu \, i \partial^2 X_\mu(z_i) + E^{\mu\nu}_i \, i \partial X_\mu(z_i) \, i \partial X_\nu(z_i) \right]
\,e^{i k_i\cdot X(z_{i})}\Bigr\rangle.
\end{aligned}
\end{equation}

The fact that taking the residue of tachyon or Yang-Mills integrand produces the correlators of higher level vertex operators can be understood as taking the OPE between two vertex operators. For example,  taking the residue of the OPE between two Yang–Mills vertex operators yields exactly the level-2 vertex operator:

\begin{equation}
\begin{split}
        &\epsilon_2\cdot i\partial X e^{ip_2X}(z_2)\epsilon_1\cdot i\partial Xe^{ip_1X}(z_1) \\
        & \sim \Bigg[ \frac{2\alpha'(\epsilon_1\cdot \epsilon_2)}{z_{1,2}^2}+ \frac{2\alpha'(\epsilon_2\cdot p_1)(\epsilon_1\cdot i\partial X)}{z_{1,2}}+(\epsilon_2\cdot i\partial X)(\epsilon_2\cdot i\partial X)\\&\hspace{2em}-\frac{4\alpha'^2(\epsilon_1\cdot p_2)(\epsilon_2\cdot p_1)}{z_{1,2}^2}-\frac{2\alpha'\left(\epsilon_1\cdot p_2)(\epsilon_2 \cdot (i\partial X+(i\partial X )^2z_{1,2})\right)}{z_{1,2}}\Bigg]\\
        &\hspace{5em}\times\Bigg[1+(p_2\cdot i\partial X) z_{1,2}+\frac{1}{2}\left((p_2\cdot  i\partial X)^2+(p_2\cdot i\partial^2 X)\right)z_{1,2}^2 \Bigg]\left(\frac{1}{z_{1,2}}\right)e^{i(p_1+p_2)X }(z_1).
\end{split}
\end{equation}
After taking the residue at \( z_{1,2} \), we obtain the same level-2 vertex operator from scaffolding procedure. The fact that the resulting higher level vertex operator naturally satisfy BRST constraint follows from the OPE expansion of two primaries of weight one contains another primary of weight one with $1/z$ singularity, i.e.
 \begin{equation}
 V(z_1)V(z_2)|_{z_1\rightarrow z_2}=\cdots+\frac{V'}{z_{1,2}}+\cdots,
 \end{equation}
where if $V$ are of weight 1, then so will $V'$.

\subsection{Zeros of Gluon amplitudes}

Recall that for the tachyon amplitude the curve-integral representation admits zeros with 
\begin{equation}\label{eq: RayZero}
\alpha'c_{i,j} = -\mathbb{N}_0, 
\quad 
\forall 
(i,j)\in\{(i,j)\mid 1\leq i\leq a-2,\;a\leq j\leq n-1\}.
\end{equation}
Consequently, the resulting Yang-Mills amplitude, which emerges from scaffolding the tachyon amplitude, must identically vanish. In terms of Yang-Mills amplitude, we simply need to identify $p_{2i}=\epsilon_i$, and $p_{2i-1}=k_i-\epsilon_i$.

The above zeros are inherited from the tachyon amplitude. As  discussed in \cite{Arkani-Hamed:2023swr}, there are emergent zeros by analyzing  conditions of scaleless integrals for the curve-integral representation of gluon amplitudes. Again starting from the $2n$-tachyon amplitude, we take the residue at poles defined by $\alpha' X_{2i-1,2i+1}= 0$, for $i = 1,2,\dots,n$.

First, we choose the triangulations involve the following internal cords:
\[
\{(1,3), (3,5),\dots,(2n-1,1)\}.
\]
The corresponding amplitude is given by:
\begin{equation}
A^{\text{Tachyon}}_{2n} = \int_0^\infty \prod_{s=1}^{n} \frac{dy_{2s-1,2s+1}}{y_{2s-1,2s+1}^2} y^{\alpha' X_{2s-1,2s+1}} \prod_{r=1}^{n-3} \frac{dy_{i_r,j_r}}{y_{i_r,j_r}^2} y^{\alpha' X_{i_r,j_r}} \prod_{i<j} F_{i,j}^{-\alpha' c_{i,j}},
\end{equation}
where $y_{i_r,j_r}$ represents the remaining internal cord variables unspecified so far. Taking the residue, the gluon amplitude can be formally expressed as:
\begin{equation}
\begin{split}
A^{\text{YM}}_{n} \equiv \int_0^\infty \prod_{t=1}^{n-3} \frac{dy_{i_t,j_t}}{y_{i_t,j_t}^2} y^{\alpha' X_{i_t,j_t}}\left(\prod_{i=1}^n \partial_{y_{2i-1,2i+1}}\right)  \left( \prod_{i<j}F_{i,j}^{-\alpha' c_{i,j}} \right)\bigg|_{y_{2a-1,2a+1}=0},
\\a\in\{1,2,\dots,n\}.
\end{split}
\end{equation}
Before analyzing the zeros, we first examine the structure of the integral. Taking the residue effectively decompose the integrand into two parts: the first is the differentiation of some function \( F_{i,j} \) while the second is setting the remaining 
$\prod_{i,j} F_{i,j}^{-\alpha' c_{i,j} - b_{i,j}} \Big|_{y_s = 0}$,
where
\(\{y_s\}=(y_{1,3},\cdots, y_{2n{-}1,1})\)
is the variable we are taking the residue on and $b_{i,j}$ some non-negative integer. For the purpose of examining scaleless integrals only the latter is relevant. 

Therefore, we would like to analyze the dependence \( F_{i,j}\big|_{y_s = 0} \) on the remaining $y_r$ variables.
To understand the behavior of \( F_{i,j} \) restricted to \( y_s = 0 \), we define a amputated graph obtained by cutting off all external legs of the original one, thus reducing the 2n-point graph to a n-point graph. Now the original pair \(\{2i-1,2i\}\) is replaced by an end-point \(\{i\}\), and the  variables \(y_{2i-1,2j-1}\) can be relabeled as \(y_{i,j}\). The corresponding \(F\)-polynomials on the amputated graph are denoted by \(F^{\rm amp}_{i,j}\). Then under the condition \( y_s = 0 \), we have the following identity:
\begin{equation}\label{eq: FIdentity}
\boxed{\quad F_{2i-1,2j-1}\bigl\lvert_{y_s=0}
=F_{2i,2j-1}\bigl\lvert_{y_s=0}
=F_{2i-1,2j}\bigl\lvert_{y_s=0}
=F_{2i,2j}\bigl\lvert_{y_s=0}
=F^{\rm amp}_{i,j}\quad}\,.
\end{equation}
This identity allows us to analyze the dependence of \( F_{i,j}\big|_{y_s = 0} \) on \( y_r\) by simply studying the $F$-polynomial of the amputated $n$-point fat-graph.

\begin{figure}[h!]
    \centering
    \includegraphics[width=0.45\linewidth]{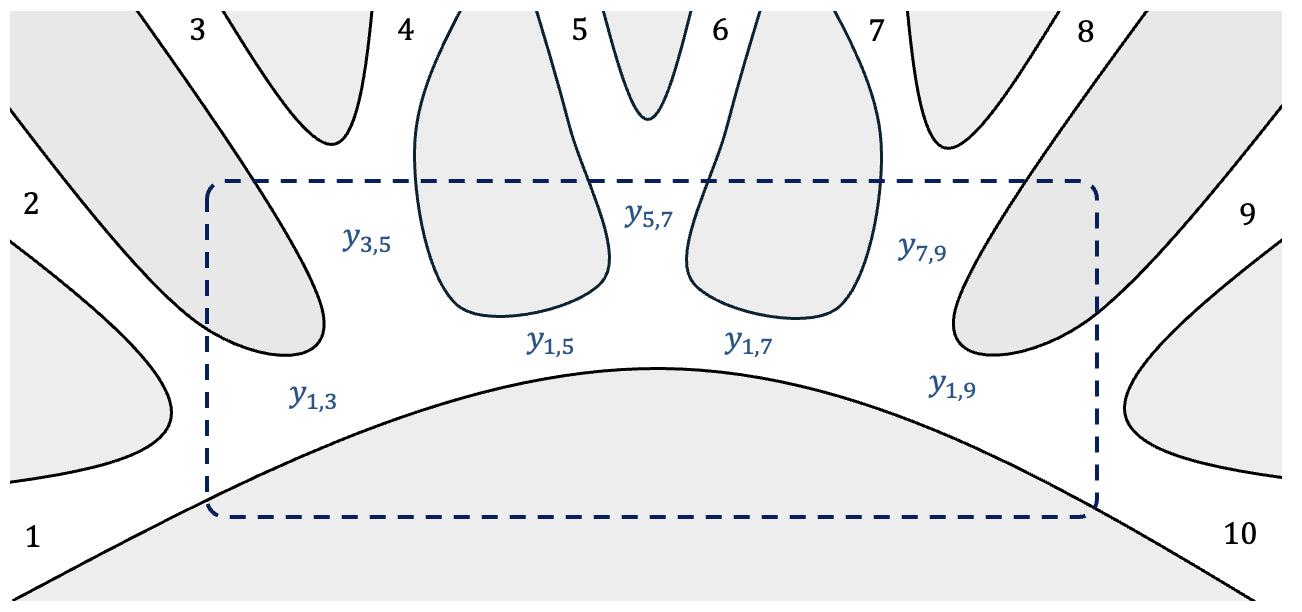} \includegraphics[width=0.45\linewidth]{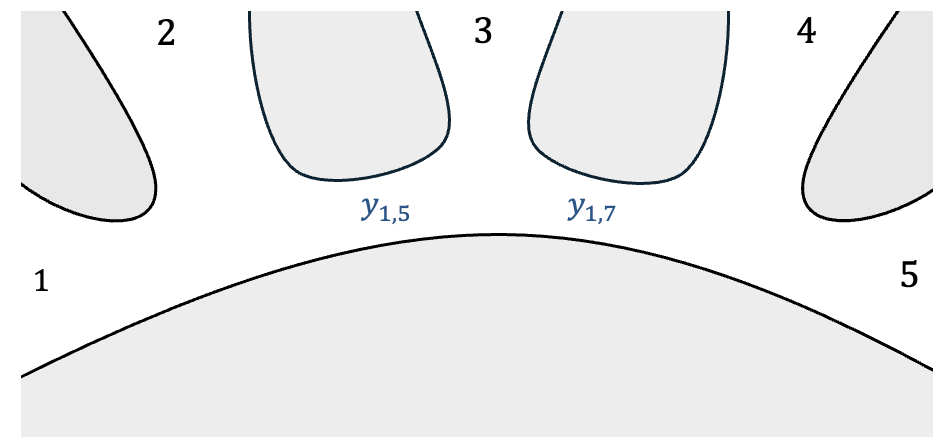}
    \caption{The $F$ polynomials obtained from scaffolding of the 10-point fat graph on the left is equivalent to the amputated 5-point fat graph with $(y_{1,3}, y_{1,4})$ of the latter identified with $(y_{1,5}, y_{1,7})$ of the former.  }
    \label{fig: 10to5}
\end{figure}

Before proceeding let us verify the identity in eq.(\ref{eq: FIdentity}) in the case of scaffolding five-point amplitudes. Starting with the following triangulation:
\[
T = \bigl\{(1,3),\,(3,5),\,(5,7),\,(7,9),\,(9,1),\,(1,5),\,(1,7)\bigr\}.
\]
Under this triangulation, the amputated five-point graph corresponds to a ray-like triangulation, where the polynomials \( F^{\rm amp}_{i,j} \) are parametrized by \( y_{1,3} \) and \( y_{1,4} \). Let us demonstrate the  correspondence between \( F_{1,7}, F_{1,8}, F_{2,7}, F_{2,8} \) and $F^{\rm amp}_{1,4}$. 
Follow the definition of $F$-polynomial we have:
\begin{align}
F_{1,7}&=(1,1)\cdot M_{R}(y_{1,3})M_{R}(y_{1,5})M_{L}(y_{1,7})M_{R}(y_{7,9})\cdot \begin{pmatrix}
1  \\
0  
\end{pmatrix}\nonumber\\
&=1+y_{1,7}+y_{1,5}y_{1,7}+y_{7,9}y_{1,7}+y_{7,9}y_{1,5}y_{1,7}+y_{1,3}y_{1,5}y_{1,7}+y_{1,3}y_{1,5}y_{1,7}y_{7,9}\nonumber\\
F_{1,8}
&=1+y_{1,7}+y_{1,5}y_{1,7}+y_{1,3}y_{1,5}y_{1,7}\nonumber\\
F_{2,7}&=1+y_{1,7}+y_{1,5}y_{1,7}+y_{1,7}y_{7,9}+y_{1,5}y_{1,7}y_{7,9}
\nonumber\\
F_{2,8}&=1+y_{1,7}+y_{1,5}y_{1,7}\,.
\end{align}
On the other hand, the corresponding $F^{\rm amp}_{1,4}$ is given by: 
\begin{align}
F^{\rm amp}_{1,4}&=1+y_{1,4}+y_{1,3}y_{1,4}=1+y_{1,7}+y_{1,5}y_{1,7},
\end{align}
where the final equality is just embedding the variables of the amputated graph in its parent. This is illustrate in fig.(\ref{fig: 10to5}). It is straightforward to see that by setting $y_s=0$ eq.(\ref{eq: FIdentity}) holds. 


Let's consider the following choice of triangulation:
\[
\mathcal{T} = \{(1,3), (3,5), \dots, (2n-1, 2n+1), (1,5), (1,7), \dots, (1, 2n-3)\}\,,
\]
in which case the amputated graph adopts a ray-like triangulation. Thus we are asking the dependence of $F^{\rm amp}_{i, j}$ on \( y_{1,2a+3} \). But this is exactly the problem that we have solved before, only $F^{\rm amp}_{i, j}$s with \((i, j) \in \{i \in \{1, 2, \dots, a\},\; j \in \{a+2, a+3, \dots, n-1\}\}\) has non-trivial dependence. Translating back to the original labeling, this corresponds to 
\((i,j) \in \{i \in \{1, 2, \dots, 2a\},\; j \in \{2a+3, 2a+4, \dots, 2n-2\}\}\). 
Thus we see that by setting
\begin{equation}
\alpha'c_{i,j} = -\mathbb{N}_0,  \quad \forall 
\label{eq: YMZero}
(i,j)\in \text{N}_{\text{YM}}
:=\bigl\{(i,j)\mid i\in\{1,2,\dots,2a\},\;j\in\{2a+3,\dots,2n-2\}\bigr\},
\end{equation}
the $y_{1,2a+3}$ dependence of the $n$-point gluon amplitude becomes scaleless, so the amplitude vanishes. As discussed in~\cite{Arkani-Hamed:2023swr}.

From the above discussion, we see that due to the identity eq.(\ref{eq: FIdentity}), any zero of the tachyon amplitude must be a zero of Yang-Mills. To recap, when scaffolding the Yang-Mills amplitude, we take the residue on the curve-integral representation of the tachyon amplitude. The resulting $F$-polynomials with kinematic exponents are simply that of the $2n$-tachyon evaluated on $y_s=0$. These are simply the $F$-polynomials of the amputated fat-graph, i.e. that of the $n$-point tachyon amplitude.  

Thus in summary, the $n$-point Yang-Mills amplitude enjoys two types of zeros, those that descends from the $2n$-point  and those of the $n$-point tachyon amplitude.

\subsection{Zeros of the Level-2 amplitudes}

\begin{figure}[h!]
    \centering
    \includegraphics[width=0.45\linewidth]{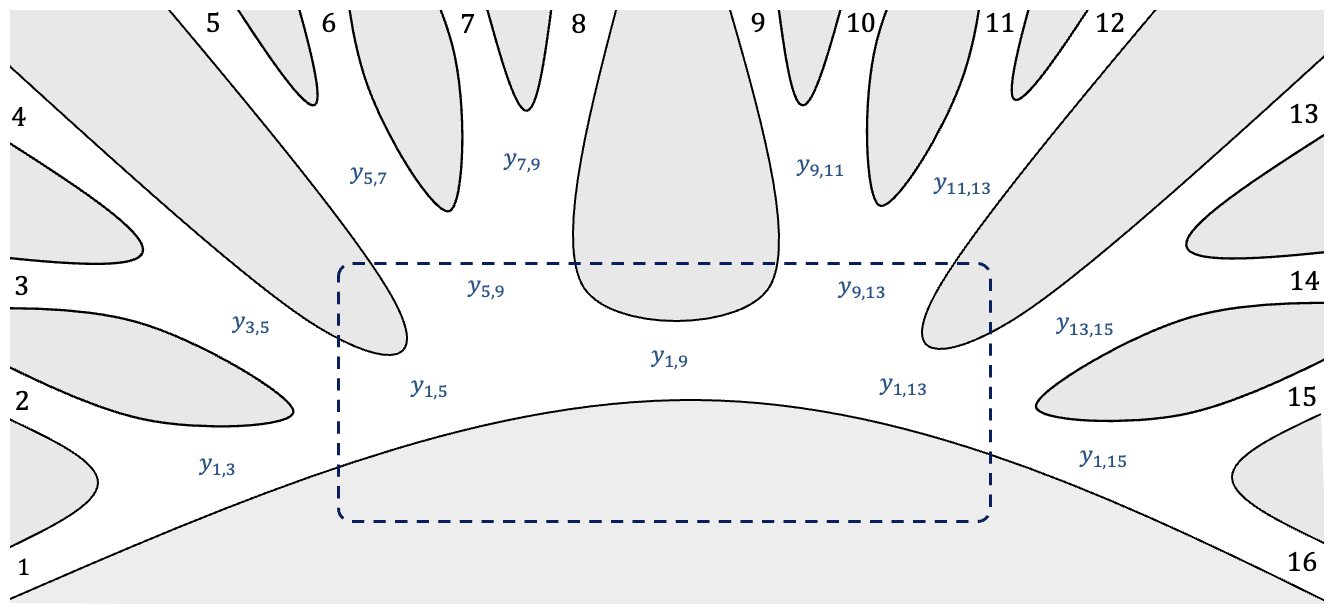} \includegraphics[width=0.45\linewidth]{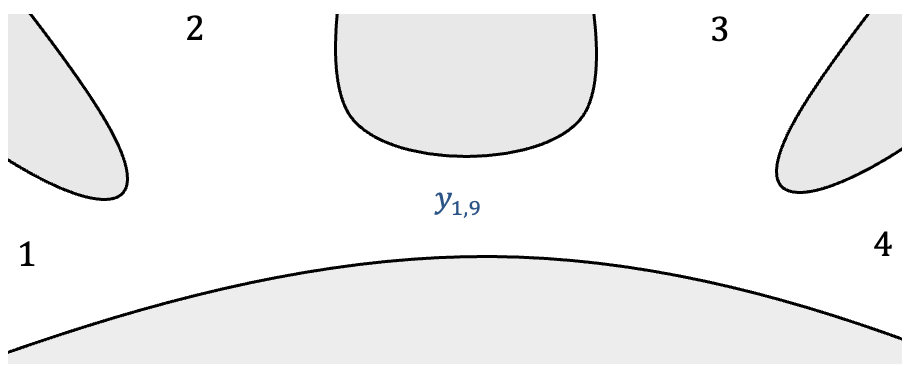}
    \caption{Double scaffolding from 16-points to 4-points. The $F$-polynomials of the parent fat-graph evaluated on $y_s=0$ reduces to the $F^{\rm amp}$ of the latter with $y_{1,3}$ identified with $y_{1,9}$ of the parent.}
    \label{fig: 16to4}
\end{figure}

Proceeding to $n$-point level-2 amplitudes, we simply apply eq.(\ref{eq: FIdentity}) twice. The amputated fat-graph is obtained by chopping off four consecutive external legs \( \{4i-3, 4i-2, 4i-1, 4i\} \) into a single effective node labeled \( \{i\} \).  We then again utilize the  relation $F_{i,j}\big|_{y_s = 0} = F^{\rm amp}_{i',j'}$ for:
\begin{align*}
i \in \{4i'{-}3, 4i'{-}2, 4i'{-}1, 4i'\},\quad
j \in \{4j'{-}3, 4j'{-}2, 4j'{-}1, 4j'\} \bigr\}.
\end{align*}
For the case where both $(i,j)\in\{4i'{-}3, 4i'{-}2, 4i'{-}1, 4i'\}$, we have  $F_{i,j}\big|_{y_s = 0}=1$. This result implies that the behavior of \( F_{i,j}\big|_{y_s = 0} \) can be deduced from the corresponding \( F^{\rm amp}_{i',j'} \) on the amputated graph, see fig.(\ref{fig: 16to4}). Then, from our ray like triangulation of the tachyon amplitude, we conclude that \( F_{i,j} \) may depend on \( y_{1, 4a+5} \) only if
\[
(i,j) \in \bigl\{ i \in \{1, 2, \dots, 4a\},\; j \in \{4a{+}5, 4a{+}6, \dots, 4n{-}4\} \bigr\}.
\]
This then implies when 
\[
\alpha'c_{i,j} \in -\mathbb{N}_0, \quad \forall 
(i,j) \in \bigl\{ i \in \{1, 2, \dots, 4a\},\; j \in \{4a{+}5, 4a{+}6, \dots, 4n{-}4\} \bigr\},
\]
the integrand becomes polynomial in \( y_{1,4a+5} \), and the integral becomes scaleless in this variable. Consequently, the $n$-point level-2 amplitude vanishes.

\paragraph{Single level-2} The zeros of mixed levels can also be straightforwardly deduced using eq.(\ref{eq: FIdentity}). Let us consider the simplest example with a level-2 vertex and three tachyons. The amplitude is given by:
\begin{equation}
\begin{split}
    &\int \frac{dz_1dz_2dz_3dz_4}{\mathrm{SL}(2,\mathbb{R})} \langle(B^\mu i\partial^2X_\mu +E^{\mu\nu}i\partial X_\mu i\partial X_\nu)e^{ik\cdot X}(z_1)e^{ip_5\cdot X}(z_2)e^{ip_6\cdot X}(z_3)e^{ip_7\cdot X}(z_4)\rangle \\
    &=\text{B}(2\alpha' k\cdot p_5-1, 2\alpha' p_5\cdot p_6+1)\left(-2\alpha' B\cdot p_5+4\alpha'^2p_5\cdot E\cdot p_5\right)\\
    &\qquad+ \text{B}(2\alpha' k\cdot p_5+1, 2\alpha' p_5\cdot p_6-1)\left(-2\alpha' B\cdot p_6+4\alpha'^2p_6\cdot E\cdot p_6\right)\\
    &\qquad+ \text{B}(2\alpha' k\cdot p_5, 2\alpha' p_5\cdot p_6)\left(8\alpha' p_5\cdot E\cdot p_6\right),
\label{23t}
\end{split}
\end{equation}
where $k=q_1+q_2$, $q_1=p_1+p_2$, $q_2=p_3+p_4$, $k^2=-\tfrac{1}{\alpha'}$, $q_i^2=0$, $p_i^2=\tfrac{1}{\alpha'}$. To determine the zero condition, we start from the 7-point tachyon amplitude with the chosen triangulation:
\[
    \mathcal{T}=\bigl\{(1,3),(3,5),(1,5),(1,6)\bigr\}.
\]
The corresponding amplitude is expressed as:
\begin{align}
A_{\text{1\textbf{L2},3\textbf{T}}}
&=\mathrm{Res}_{\alpha' X_{1,3}}\mathrm{Res}_{\alpha' X_{3,5}}\mathrm{Res}_{\alpha' X_{1,5}=-1}\,A^{\text{Tachyon}}_{7}\nonumber \\
&=\mathrm{Res}_{\alpha' X_{1,3}}\mathrm{Res}_{\alpha' X_{3,5}}\mathrm{Res}_{\alpha' X_{1,5}=-1}\nonumber\\&\qquad\times\Biggl[\,
\int \biggl( \frac{dy_{1,3}}{y^2_{1,3}}\,y_{1,3}^{\,\alpha' X_{1,3}}\frac{dy_{3,5}}{y^2_{3,5}}\,y_{3,5}^{\,\alpha' X_{3,5}}\frac{dy_{1,5}}{y^2_{1,5}}\,y_{1,5}^{\,\alpha' X_{1,5}}\biggr)
\frac{dy_{1,6}}{y^2_{1,6}}\,y_{1,6}^{\,\alpha' X_{1,6}}\biggr)
\!\prod_{i<j} F_{i,j}^{-\alpha' c_{i,j}}
\Biggr]\,\nonumber\\
&=\int\frac{dy_{1,6}}{y^2_{1,6}}\,y_{1,6}^{\,\alpha' X_{1,6}}\left(\partial_{y_{1,3}}\partial_{y_{3,5}}\partial_{y_{1,5}}\right)  \left( \prod_{i<j}F_{i,j}^{-\alpha' c_{i,j}} \right)\bigg|_{y_{1,3}=y_{3,5}=y_{1,5}=0}.
\end{align}
We then determine under which conditions the integral becomes scaleless, implying that the amplitude vanishes. The analysis is similar to the previous discussion and using the small graph to help us determine the dependence of $F_{i,j}$ on $y_{1,6}$, so we omit the details and state the conclusion: only $F_{ij}$ terms with indices $(i, j)$   
\[
      (i,j)\in \text{N},
      \quad\text{where}\quad
      \text{N} = \bigl\{(i,j)\mid i=1,2,3,4 ,\; j=6\bigr\}
\]
can involve the chosen variable $y_{1,6}$ after taking all scaffolding residues. Therefore, when we set $c_{i,j}=0$, $\forall(i,j)\in \text{N}$, the integral becomes scaleless, and the amplitude vanishes. This is equivalent to the following condition:
\[
B(q_1,q_2,\epsilon_1, \epsilon_2)\cdot p_6 =E^\mu(q_1,q_2,\epsilon_1, \epsilon_2) \cdot  p_6=0,
\]
given $p_2=\epsilon_1$, $p_4=\epsilon_2$, and $q_i \cdot \epsilon_i=0$. Plugging this condition into the amplitude in eq.(\ref{23t}), we see that the second and last terms vanish identically, while the first term vanishes due to the denominator of the beta function yields:
\[
 \frac{1}{\Gamma(2\alpha' k \cdot p_5 + 2\alpha' p_5 \cdot p_6)} = \frac{1}{\Gamma(-1)}=0\,.
\]

\section{Super-string I: Super-tachyon seed}

We now consider super-string amplitudes in its curve-integral representation. Similarly to the bosonic string, it is useful to begin with the super-tachyon amplitude. We will need the \( q = -1 \) and \( q = 0 \) super-tachyon vertex operators, where \( q \) refers to the superghost charge (i.e., the picture number):
\begin{equation}
    V_\text{ST}^{-1}(p,z) = e^{-\phi} e^{ip\cdot X} (z),\quad V_\text{ST}^{0}(p,z) = p\cdot\psi e^{ip\cdot X} (z)\,,
\end{equation}
where $p^2 = \frac{1}{2\alpha'}$. The \( 2n \)-point super-tachyon scattering amplitude is then given by:
\begin{equation}
A^{\text{ST}}_{2n} 
= \int \frac{\prod_{a=1}^{2n} dz_{2a}}{\mathrm{SL}(2,\mathbb{R})}\,\Bigl\langle\prod_{a=1}^{2n} V_\text{ST}^{q_a}(z_a)\Bigr\rangle,
\end{equation}
where $q_a$ satisfy $\sum_aq_a=-2$ by superghost charge conservation. 
\subsection{Curve-Integral representation and its zeros}

It is useful to introduce grassmann-odd variables $\theta$ for the $0$-picture operators such that:
\begin{align}
A^{\text{ST}}_{2n} 
&=\int \frac{ d^{2n}zd^{2n-2}\theta}{\mathrm{SL}(2,\mathbb{R})}\,\Bigl\langle\prod_{a=1}^{2n\setminus{i,j}} e^{ip_a\cdot X(z_a)+\theta_a p_a\psi(z_a)}\,e^{-\phi(z_i)}e^{ip_i\cdot X(z_i)}\,e^{-\phi(z_j)}e^{ip_j\cdot X(z_j)}\Bigr\rangle\nonumber\\
&= \int\frac{d^{2n}z\,d^{2n-2}\theta}{\mathrm{SL}(2,\mathbb{R})}\frac{(-1)^{i+j+1}}{z_{i,j}} \prod_{a<b}\left|z_a - z_b + \theta_a\theta_b\right|^{-\alpha' c_{a,b}}\nonumber\\
&=\int\frac{d^{2n}z}{\mathrm{SL}(2,\mathbb{R})}\frac{(-1)^{i+j+1}}{z_{i,j}}\prod_{a<b}z_{a,b}^{-\alpha'c_{a,b}}\text{Pf}(D_{a,b}^{i,j}),
\end{align}
where $D_{a,b}^{i,j}$ is the reduced pfaffian on the skew-symmetric matrix $D_{a,b}$ with columns and rows $i,j$ reomoved. The matrix $D_{a,b}$ is defined as:
\begin{equation}
D_{a,b} = \begin{cases}
\frac{-\alpha' c_{a,b}}{z_{a,b}}, & a<b\\[6pt]
-D_{b,a}, & a>b
\end{cases}.
\end{equation}
The independence of the choice of $i,j$ is a reflection of picture independence. As in the bosonic string, using the identities in eq.(\ref{eq: CurveId}), the amplitude becomes:
\begin{align}\label{eq: SuperT}
A^{\text{ST}}_{2n} 
&= \int \frac{d^{2n}z}{\mathrm{SL}(2,\mathbb{R})} 
\prod_{i=1}^{2n} z_{i,i+1}^{-1}\prod_{j=1}^{2n} z_{j,j+2}^{1/2}  \prod_C u_C^{\alpha' X_C} \frac{1}{z_{a,b}} \mathrm{Pf}(D^{a,b})\nonumber\\
&= \int \prod_{d=1}^{2n-3} \frac{dy_{i_d,j_d}}{y_{i_d,j_d}} \prod_C u_C^{\alpha' X_C} \prod_{i=1}^{2n} z_{i,i+2}^{\frac{1}{2}} \cdot \frac{1}{z_{a,b}} \mathrm{Pf}(D^{a,b}).
\end{align}
Pfaffian of $D^{a,b}$ is the summation over $\Pi$ the set of all partitions of \(\{1,2,\dots,2n\}\setminus{\{a,b\}}\) into pairs without regard to order.
An element \(\alpha\in\Pi\) can be written as:
\[
\alpha=\bigl\{(i_1,j_1),(i_2,j_2),\ldots,(i_n,j_n)\bigr\},
\]
with \(i_k<j_k\) and \(i_1<i_2<\cdots<i_n\).  
Let $\pi_\alpha$ be defined as:
\[
\pi_\alpha=
\begin{bmatrix}
1 & 2 & 3 & 4 & \hat{a}&\cdots&\hat{b}&\cdots & 2n-1 & 2n\\
i_1 & j_1 & i_2 & j_2 & \cdots& \cdots& \cdots& \cdots & i_n & j_n
\end{bmatrix},
\]
which is the corresponding permutation.  Given a partition \(\alpha\) as above, define $D^{a,b}_\alpha$:
\[
D^{a,b}_\alpha
=\operatorname{sgn}(\pi_\alpha)\,
D_{i_1,j_1}\,D_{i_2,j_2}\,\cdots\,D_{i_n,j_n}.
\]
The Pfaffian of \(D^{a,b}\) is then given by
\begin{equation}
\operatorname{Pf}(D^{a,b})=\sum_{\pi_\alpha\in\Pi }\operatorname{sgn}(\pi_\alpha)\ \prod_{p \in \alpha} \frac{-\alpha' c_p}{z_p} \,.
\end{equation}
Thus we have:
\begin{align}\label{eq: superTachyon}
A^{\text{ST}}_{2n}&= \int \prod_{d=1}^{2n-3} \frac{dy_{i_d,j_d}}{y_{i_d,j_d}} \prod_C u_C^{\alpha' X_C} \left[\prod_{i=1}^{2n} z_{i,i+2}^{\frac{1}{2}} \cdot \frac{1}{z_{a,b}} \sum_{\pi_\alpha\in\Pi }\operatorname{sgn}(\pi_\alpha)\ \prod_{p \in \alpha} \frac{-\alpha' c_p}{z_p} \right]\nonumber\\
&= \int \prod_{d=1}^{2n-3} \frac{dy_{i_d,j_d}}{y_{i_d,j_d}}
   \prod_{C} u_C^{\alpha' X_C}
   \left[
     \sum_{\pi_\alpha\in\Pi}\!
       \operatorname{sgn}(\pi_\alpha)\,
       \prod_{p\in\alpha}(-\alpha' c_p)\,
       \prod_{C}u_{C}^{\,n^{\pi_\alpha}_{C}}
   \right].
\end{align}
According to the analysis in the appendix of~\cite{Cao:2025lzv}, the last exponent is a half-integer that lies in the range:
\[
n^{\pi_\alpha}_{i_d,j_d} \in \left[-1,\; \frac{j_d - i_d}{2} - 1\right].
\]
Using eq.(\ref{eq:u_from_F_T}) and (\ref{eq:u_from_z}), we finally arrive at:
\begin{align}\label{eq: SuperTC}
A^{\text{ST}}_{2n}&= \int \prod_{d=1}^{2n-3} \frac{dy_{i_d,j_d}}{y_{i_d,j_d}} y_{i_d,j_d}^{\alpha' X_{i_d,j_d}} \prod_{i<j} F_{i,j}^{-\alpha' c_{ij}} \left(\sum_{\pi_\alpha\in\Pi }\operatorname{sgn}(\pi_\alpha)\frac{1}{F_{a,b}}\prod_{p \in \alpha} \frac{-\alpha' c_p}{F_p}\prod_{d=1}^{2n-3} y_{i_d,j_d}^{n^{\pi_\alpha}_{i_d,j_d}} \right).
\end{align}

We can now analyze the zeros of the super-tachyon amplitude. There are now two types, those that reflect the vanishing of the Pfaffian in eq.(\ref{eq: SuperT}), and the emergence of our familiar scaleless integrals. Let's begin with the former.

\paragraph{Zeros in the Pfaffian.} The Pfaffian vanishes when ever the rank is lowered. This can be achieved when we set certain  \(c_{i,j} = 0\). As an example, consider the Pfaffain for $6$-point amplitude. The reduced Pfaffain \(D^{3,5}\) will vanish when
\begin{equation}\label{eq: PfaffianZero}
c_{2,4} \;=\; c_{2,6} \;=\; c_{4,6} \;=\; 0\,.
\end{equation}
This has a natural generalization to \(2n\)-point. Consider the reduced Pfaffian \(D^{2n-3,2n-1}\).
If we set:
\begin{equation}
\label{SUSY0}
c_{i,j} = 0,\quad\text{ when i, j are both even},
\end{equation}then, by a suitable permutation, the matrix \(D^{2n-3,2n-1}\) can be arranged into a block form whose lower-right \(n\times n\) block is identically zero. In particular, focusing on the last \(n\) column vectors, one sees that their span lies entirely within the first \(n-2\) dimensions, so there is a linear dependence. Hence \(D^{2n-3,2n-1}_{2n}\) is not full rank, and its Pfaffian vanishes under these conditions.

Similarly, setting all \(c_{i,j} = 0\) for odd indices \(i,j\) leads to the same conclusion: the Pfaffian vanishes due to an analogous argument.

\paragraph{Zeros from scaleless integrals} Recall that for the bosonic tachyon amplitude, the amplitude vanishes for the kinematic configurations in eq.(\ref{eq: RayZero}). Since the only difference between the bosonic and super-tachyon amplitude is the terms in the parenthesis in eq.(\ref{eq: SuperTC}), we will show that each term indeed vanishes. Denote the pairs $(i,j)$ identified in eq.(\ref{eq: RayZero}) as $\text{N}_d$.
It is important to note that each term in the integrand contains a factor of \(\frac{1}{F_{a,b}}\), which determined by the picture choice. In order to ensure that the integration over \(y_{1,d}\) becomes scaleless, we must carefully select the picture \((a,b)\) such that the corresponding \(F_{a,b}\) does not depend on \(y_{1,d}\).
To this end, we choose the picture \((a,b)\) satisfying \((a,b) \notin \text{N}_d\). This guarantees that the factor \(F_{a,b}\) will not obstruct the scalelessness of the integral on $y_{1,d}$
We then classify the terms according to whether the partition \(\alpha\) intersects \(\text{N}_d\):
\begin{enumerate}
  \item If \(\alpha \cap \text{N}_d \neq \emptyset\), then pick any \(p\in \alpha\cap \text{N}_d\):
    \begin{enumerate}
      \item If \(c_p = 0\), the term vanishes immediately.
      \item If \(\alpha'c_p \in -\mathbb{N}_0\), then the exponent of \(F_p\) is a nonnegative integer, making the \(y_{1,d}\) integral scaleless and thus the term vanishes.
    \end{enumerate}
  \item If \(\alpha \cap \text{N}_d = \emptyset\), then exactly as in the bosonic case the integral degenerates to a scaleless form and the term vanishes.
\end{enumerate}

In summary, we see that the super-tachyon enjoys the same set of zeros as the bosonic tachyon amplitude, with additional zeros from the pffafian structure. The latter new set of zeros can be viewed as a reflection of the underlying supersymmetry. Indeed as we will see shortly, the zeros inherited by super Yang-Mills will exactly kill the non-susy operator.

\subsection{Scaffolding Super Yang-Mills and its zeros}
\label{yangmillsvertexscaffold}
Equipped with the super-tachyon amplitude, we can directly scaffold super Yang-Mills amplitude. Once again, that this gives the correct amplitude can be understood from the fact that taking the residue for the OPE of two tachyon vertex operators generate the Yang-Mills vertex operator. Once again we first identify the momenta of the two operators $p_{1} + p_{2} = k$, $p_2=\epsilon$ and $c_{1,2}=\tfrac{1}{\alpha'}$.\footnote{The factor of 2 difference compared to bosonic tachyon amplitude is due to the tachyon mass in superstring has $p^2=\frac{1}{2\alpha'}$}

There are three possible combinations of these two types of super-tachyon vertex operators, namely: \((-1, -1)\), \((0, -1)\), and \((0, 0)\). We will show that their merging yields Yang–Mills vertex operators in picture \(-2\), \(-1\), and \(0\), respectively.

\paragraph{Case 1: {Super-tachyon \((-1, -1) \rightarrow V_{\rm{SYM}}^{-2}\)}} 

\begin{equation}
\begin{split}
    &V_\text{ST}^{-1}(p_2,z_2)V_\text{ST}^{-1}(p_1,z_1)=(e^{-\ln(z_2-z_1)})(e^{-\phi(z_2)}e^{-\phi(z_1)})(e^{2\alpha'p_1\cdot p_2 \ln(z_2-z_1)})(e^{ip_2\cdot X(z_2)}e^{ip_1\cdot X(z_1)})\\
    &\sim e^{{-}\ln z_{1,2}}(e^{{-}\phi(z_1)}{-}z_{1,2}\partial \phi e^{{-}\phi(z_1)} )e^{{-}\phi(z_1)}e^{{-}\ln z_{1,2}}(e^{ip_2\cdot X(z_1 )}{+}z_{1,2}p_2 \cdot i\partial X e^{ip_2\cdot X(z_1 )}) e^{ip_1\cdot X(z_1)}\\
    &\sim  \frac{1}{z^2_{1,2}}\left(1+p_2\cdot\partial iX z_{1,2}\right)\left(1-\partial\phi z_{1,2} \right) e^{-2\phi} e^{ik\cdot X}(z_1)\,,
\end{split}
\end{equation}
where in the second line we expand $z_2$ near $z_1$.  Taking the residue yields a vertex operator of the form:
\begin{equation}
    V_{\text{SYM}}^{-2}=(\epsilon \cdot i\partial X -\partial\phi) e^{-2\phi}e^{ik\cdot X}.
\end{equation}

\paragraph{Case 2: {Super-tachyon \((0, -1) \rightarrow V_{\rm{SYM}}^{-1}\)}} 

\begin{equation}
\begin{split}
    &V_{\rm{ST}}^{0}(p_2,z_2)V_{\rm{ST}}^{-1}(p_1,z_1) = p_2\cdot\psi e^{ip_2\cdot X} (z_2)\,\,e^{-\phi} e^{ip_1\cdot X} (z_1)   \\
    &\sim \left( \frac{1}{z_{1,2}}\right)\left(p_2\cdot \psi\right)\left(1+p_2\cdot i\partial X z_{1,2}\right) e^{-\phi} e^{ik\cdot X}(z_1)
\end{split}
\end{equation}
Taking the residue yields a vertex operator of the form:
\begin{equation}
    V_{\text{SYM}}^{-1}=(\epsilon \cdot \psi) e^{-\phi}e^{ik\cdot X}.
\end{equation}

\paragraph{Case 3: {Super-tachyon \((0, 0) \rightarrow V_{\rm{SYM}}^{0}\)}} 

\begin{equation}
\begin{split}
    &V_{\text{ST}}^{0}(p_2,z_2)V_{\text{ST}}^{0}(p_1,z_1) = p_2\cdot\psi e^{ip_2\cdot X} (z_2)\,\,p_1\cdot \psi e^{ip_1\cdot X} (z_1)    \\
    &\sim \left( \frac{1}{z_{1,2}}\right)\left(\frac{p_1\cdot p_2}{z_{1,2}}+(p_2\cdot \psi)(p_1\cdot\psi)\right)\left(1+p_2\cdot i\partial X z_{1,2}\right) e^{ik\cdot X}(z_1)
\end{split}
\end{equation}
Taking the residue yields a vertex operator of the form:
\begin{equation}
    ((p_1\cdot p_2)(\epsilon\cdot i\partial X)+(\epsilon\cdot\psi)((k-\epsilon)\cdot\psi)) e^{ik\cdot X}=(-\frac{1}{2\alpha'}(\epsilon\cdot i\partial X)+(\epsilon\cdot\psi)(k\cdot\psi)) e^{ik\cdot X}
\notag
\end{equation}
\begin{equation}
    \Longrightarrow \quad V_{\text{SYM}}^{0}=((\epsilon\cdot i\partial X)+2\alpha'(k\cdot\psi)(\epsilon\cdot\psi)) e^{ik\cdot X},
\end{equation}
up to an overall factor. Thus by taking the residue of the super-tachyon amplitude, we immediately land on the $n$-point Yang-Mills amplitude, 
\begin{equation}
A^{\text{SYM}}_{n} 
= \int \frac{\prod_{i=1}^n dz_{i}}{\mathrm{SL}(2,\mathbb{R})}\,\Bigl\langle\prod_{i=1}^{n} V_{\text{SYM}}^{q_i}(z_i)\Bigr\rangle,
\end{equation}
where $q_a$ satisfy $\sum_iq_i=-2$. In appendix~\ref{sec: App} we scaffold directly on the correlator and demonstrate that the above results is reproduced. 

Explicitly, the curve representation of SYM amplitude can be written as:
\begin{align}
A^{\text{SYM}}_{n} 
&=\prod_{i=1}^n \mathrm{Res}_{\alpha'X_{2i{-}1,2i{+}1}}\,A^{\text{ST}}_{2n}\nonumber\\
&=\prod_{i=1}^n \mathrm{Res}_{\alpha'X_{2i{-}1,2i{+}1}}\Biggl[\,
\int \biggl(\prod_{s=1}^n \frac{dy_{2s-1,2s+1}}{y_{2s-1,2s+1}}\,y_{2s-1,2s+1}^{\,\alpha' X_{2s-1,2s+1}}\biggr)
\!\biggl(\prod_{r=1}^{n-3} \frac{dy_{i_r,j_r}}{y_{i_r,j_r}}\,y_{i_r,j_r}^{\,\alpha' X_{i_r,j_r}}\biggr)
\!\prod_{i<j} F_{i,j}^{-\alpha' c_{i,j}}\nonumber\\
&\qquad\quad\times
\Bigl(\sum_{\pi_\alpha\in\Pi} 
\operatorname{sgn}(\pi_\alpha)\,\frac{1}{F_{a,b}}\,
\prod_{p\in \alpha}\frac{c_p}{F_p}\,
\prod_{s=1}^n y_{2s-1,2s+1}^{\,n^{\pi_\alpha}_{2s-1,2s+1}}\,
\prod_{r=1}^{n-3}y_{i_r,j_r}^{\,n^{\pi_\alpha}_{i_r,j_r}}
\Bigr)\Biggr]\,.
\end{align}

We now turn to the vanishing of the SYM amplitude. We have already seen that zeros emerging from ray-like triangulation is shared between the tachyon and super-tachyon   amplitude. We now show that zeros from the scaffolding triangulation, eq.(\ref{eq: YMZero}) is also shared by the SYM amplitude.

First, recall the result from the analysis of zeros in the Yang-Mills amplitude: the functions \( F_{i,j} \big|_{y_s = 0} \) can depend on \( y_{1,2a+3} \) only when
\[
(i,j)\in \text{N}_{\text{YM}} \;:=\; \bigl\{(i,j)\mid 1 \le i \le 2a,\; 2a+3 \le j \le 2n-2 \bigr\}\,.
\]

Moreover, we know that the only potential obstruction to the integral being scaleless comes from the exponents of these \( F_{i,j} \) terms. If the exponent is not a non-negative integer, the corresponding integral may fail to be scaleless. Thus, we must carefully analyze the exponents of these \( F \)-polynomials.

Then as in the super-tachyon case, we note the presence of a prefactor \( \frac{1}{F_{a,b}} \) in each term. For consistency and convenience, we again choose the picture such that \( (a,b) \notin \text{N}_{\text{YM}} \). This ensures that the factor \( F_{a,b} \) does not interfere with the scalelessness of the integration over \( y_{1,2a+3} \).

We are now ready to examine whether setting \( \alpha'c_{i,j} = -\mathbb{N}_0 \), $\forall$ \( (i,j) \in \text{N}_{\text{YM}} \) ensures the vanishing of the SYM amplitude. We claim that under this choice, every term in the integrand vanishes, and therefore the full amplitude must vanish. To demonstrate this, we split the Pfaffian partitions \( \alpha \) into two cases:
\begin{enumerate}
  \item If \( \alpha \cap \text{N}_{\text{YM}} \neq \emptyset \), then consider all pairs \( (i,j) \in \alpha \cap \text{N}_{\text{YM}} \), and denote them as \( (i_{s_1}, j_{s_1}), (i_{s_2}, j_{s_2}), \dots, (i_{s_k}, j_{s_k}) \).  
  We further divide the discussion into two cases, depending on whether any of the corresponding coefficients \( c_{i_{s_h}, j_{s_h}} \) vanishes:
  
  \begin{enumerate}
    \item If there exists some \( h \) such that \( c_{i_{s_h}, j_{s_h}} = 0 \), then the corresponding term in the integrand is proportional to \( c_{i_{s_h}, j_{s_h}} \), and hence vanishes.
    
    \item Otherwise, all \( \alpha'c_{i_{s_h}, j_{s_h}} \) are negative integers. In this case, before taking the residue, the exponent of each corresponding \( F_{i_{s_h}, j_{s_h}} \) is
    \[
    -\alpha'c_{i_{s_h}, j_{s_h}} - 1 \in \mathbb{N}_0,
    \]
    which is still a non-negative integer. Therefore, exponents of all such terms $F_{i,j}$ for $(i,j)\in \text{N}_{\text{YM}}$ are nonnegative integer. Then follow the similar process as in the YM cases, the integral still becomes scaleless after taking the residue. Thus, these terms also vanish.
  \end{enumerate}

  \item If \( \alpha \cap \text{N}_{\text{YM}} = \emptyset \), then the product \( \prod_{(i,j)\in \text{N}_{\text{YM}}} F_{i,j}^{-\alpha' c_{i,j}} \) is a polynomial in \( y_{1,2a+3} \), since all \( \alpha'c_{i,j} \in -\mathbb{N}_0 \). As in the bosonic case, taking the residue either produces a polynomial factor in \( y_{1,2a+3} \) or imposes the constraint \( y_{2a-1,2a+1} = 0 \). In both situations, the resulting integral over \( y_{1,2a+3} \) is scaleless, and hence these terms also vanish.
\end{enumerate}

In summary, we classified the integrand terms before taking the residue into two categories, depending on whether the Pfaffian partition \(\alpha\) intersects the set \(\text{N}_{\text{YM}}\). For each case, we analyzed the structure of the resulting terms after performing the scaffolding residue. In both categories, we found that each term becomes a scaleless integral with respect to \( y_{1,2a+3} \), provided the following condition holds:
\[\alpha'
c_{i,j} \in -\mathbb{N}_0, \quad \forall (i,j) \in \text{N}_{\text{YM}}.
\]
Under this condition, all such integrals vanish, implying the vanishing of the SYM amplitude.


Beyond the zeros shared with bosonic Yang-Mills, there are additional zeros as a reflection of supersymmetry. Recall that the super-tachyon amplitude vanishes for special configurations. For example, at six points, the amplitude vanishes for $c_{2,4} =c_{1,6}=c_{4,6}=0$. Upon scaffolding this implies that the three-point super Yang-Mills must vanish under:
\[
\epsilon_{1}\!\cdot\!\epsilon_{2}
\;=\;\epsilon_{1}\!\cdot\!\epsilon_{3}
\;=\;\epsilon_{2}\!\cdot\!\epsilon_{3}
\;=\;0.
\]
This set of zeros project out the \(F^{3}\) correction in bosonic string Yang-Mills amplitude. Similarly the scaffolding image of eq.(\ref{SUSY0}) is that the amplitude vanishes when \ 
\[
\epsilon_i \!\cdot\!\epsilon_j = 0,\quad \forall\,1 \le i < j \le n.
\]
Indeed the four-point bosonic Yang-Mills amplitude can be written as $B\frac{\Gamma(1{-}s)\Gamma(1{-}t)}{\Gamma(1+u)}$~\cite{Huang:2016tag}, where 
\begin{eqnarray}
\! \! \! \! \! \! \! \! \! \! B= A_{\rm YM} + (2 \alpha')^2 s_{13} \left[ \left( 
{ f_{12} f_{34} \over s_{12}^2 (1 - s_{12}) } + \textrm{cyc}(2,3,4) \right) - 
{ g_1 g_2 g_3 g_4 \over s_{12}^2 s_{13}^2 s_{14}^2 } \right] \, ,
\notag
\end{eqnarray}
with gauge invariant constituents $f_{ij} \equiv (\epsilon_i \cdot  \epsilon_j)  (k_i \cdot k_j)- (k_i \cdot \epsilon_j) ( k_j \cdot \epsilon_i)$ and $ g_i \equiv (k_{i{-}1} \cdot \epsilon_i)  s_{i , i{+}1} {-} (k_{i{+}1} \cdot \epsilon_i) s_{i{-}1, i}$. Our zero condition forces the square bracket to be zero.

\section{Super-string II: Gluino seed}
In this section we derive the curve-integral representation for pure gluino amplitudes. The curcial step is to organize the world-sheet integrand into manifestly $\text{SL(2,\,}\mathbb{R})$ invariant blocks. We present the result at 4- and 6-point amplitudes. We leave the general discussion to future work.

\subsection{Curve-representation and its zeros}
Let's begin with the gluino vertex operators: 
 $V_{\text{Gluino}}^{-1/2}$ and $V_{\text{Gluino}}^{1/2}$, 
\begin{alignat}{3}
&V_{\text{Gluino}}^{-1/2}(u,p,z)\, &&=\, u^\alpha \Theta_\alpha e^{-\phi/2} e^{ip\cdot X}(z) \nonumber \\
&V_{\text{Gluino}}^{1/2}(u,p,z)\, &&=\, -2\left[ Q_\text{BRST},\, \xi(z)V_{\text{Gluino}}^{-1/2}(u,p,z) \right] \nonumber \\
&                     &&=\, \frac{1}{2\sqrt{\alpha'}} u^\alpha \left( i \partial X_\mu + \frac{\alpha'}{2} k_\nu \psi^\nu \psi_\mu \right) (\gamma^\mu \Theta)_\alpha e^{\phi/2} e^{ip\cdot X}(z),
\end{alignat}
where \( \Theta_\alpha \) is the left-handed spin field, and \( u^\alpha \) is a 10-dimensional Majorana-Weyl spinor satisfying the Dirac equation \( u\slashed{p}  = 0 \), with \( p_i^2 = 0 \).
\subsubsection*{Four-Point Gluino Amplitude}
The four-point correlator is given by:
\begin{equation}
\begin{split}
    &\Bigl\langle V_{\text{Gluino}}^{-1/2}(z_1)V_{\text{Gluino}}^{-1/2}(z_2)V_{\text{Gluino}}^{-1/2}(z_3)V_{\text{Gluino}}^{-1/2}(z_4)\Bigr\rangle\\
    &=u_\alpha u_\beta u_\lambda u_\delta \left( \frac{(\gamma^\mu C)_{\alpha\beta}(\gamma_\mu C)_{\lambda\delta}}{2z_{1,2}z_{2,3}z_{2,4}z_{3,4}} +\frac{(\gamma^\mu C)_{\alpha\lambda}(\gamma_\mu C)_{\delta\beta}}{2z_{1,3}z_{3,4}z_{3,2}z_{4,2}}+\frac{(\gamma^\mu C)_{\alpha\delta}(\gamma_\mu C)_{\beta\lambda}}{2z_{1,4}z_{4,2}z_{4,3}z_{2,3}}\right)\prod_{1\le i<j\le 4}z_{i,j}^{-\alpha'c_{i,j}}\\
    &=u_\alpha u_\beta u_\lambda u_\delta\left( \frac{(\gamma^\mu C)_{\alpha\beta}(\gamma_\mu C)_{\lambda\delta}}{2z_{1,2}z_{2,3}z_{3,4}z_{1,4}}+\frac{(\gamma^\mu C)_{\alpha\lambda}(\gamma_\mu C)_{\delta\beta}}{2z_{1,3}z_{1,4}z_{2,3}z_{2,4}}\right)\prod_{1\le i<j\le 4}z_{i,j}^{-\alpha'c_{i,j}},
\end{split}
\end{equation}
where we have used the identity $(\gamma^\mu C)_{\alpha\beta}(\gamma_\mu C)_{\lambda\delta}+(\gamma^\mu C)_{\alpha\lambda}(\gamma_\mu C)_{\delta\beta}+(\gamma^\mu C)_{\alpha\delta}(\gamma_\mu C)_{\beta\lambda}=0$. The amplitude can now be converted into curve-integral form by forming cross ratios, which are $\text{SL(2,\,}\mathbb{R})$ invariants:
\begin{equation}
\begin{split}
    A^{\text{Gluino}}_{4}&=u_\alpha u_\beta u_\lambda u_\delta\int\frac{d^4z}{\text{SL(2,\,}\mathbb{R})\mathbb{PT}}
    \prod_{i<j}z_{i,j}^{-\alpha'c_{i,j}}\mathbb{PT}\left( \frac{(\gamma^\mu C)_{\alpha\beta}(\gamma_\mu C)_{\lambda\delta}}{2z_{1,2}z_{2,3}z_{3,4}z_{1,4}}+\frac{(\gamma^\mu C)_{\alpha\lambda}(\gamma_\mu C)_{\delta\beta}}{2z_{1,3}z_{1,4}z_{2,3}z_{2,4}}\right)\\ 
    &= u_\alpha u_\beta u_\lambda u_\delta  \int \frac{dy_{1,3}}{y_{1,3}} \prod_C u_C^{\alpha'X_C} 
    \left(\frac{-(\gamma^\mu C)_{\alpha\beta}(\gamma_\mu C)_{\lambda\delta}}{2}+\frac{-(\gamma^\mu C)_{\alpha\lambda}(\gamma_\mu C)_{\delta\beta}}{2} \,u_{1,3} \right)\\
    &=u_\alpha u_\beta u_\lambda u_\delta  \int \frac{dy_{1,3}}{y_{1,3}} y_{1,3}^{\alpha'X_{1,3}} \prod_{i<j}F_{i,j}^{-\alpha'c_{i,j}} 
    \left( \frac{-(\gamma^\mu C)_{\alpha\beta}(\gamma_\mu C)_{\lambda\delta}}{2}+\frac{-(\gamma^\mu C)_{\alpha\lambda}(\gamma_\mu C)_{\delta\beta}}{2} \,y_{1,3}\frac{F_{1,2}F_{3,4}}{F_{1,3}F_{2,4}}\right)
\end{split}
\end{equation}
We now consider the zeros of this amplitude. Since only \(F_{1,3}\) can depend on \(y_{1,3}\), we focus on its exponent. In the first term the exponent of \(F_{1,3}\) is $-\alpha'c_{1,3}$
and in the second term it is
$
-\alpha'c_{1,3}-1$. Hence the four-point amplitude exhibits zeros when 
\begin{equation}
\alpha'c_{1,3}=-\mathbb{N}.
\end{equation}

\subsubsection*{Six-Point Gluino Amplitude}
In the six-point gluino amplitude now involves one $V_{\text{Gluino}}^{1/2}$. We consider: 
\begin{align}
\label{eq:I6_definition}
A^{\text{Gluino}}_{6}&=\int\frac{d^6z}{SL(2,\mathbb{R})}\langle V_F^{-1/2}(z_1)V_{\text{Gluino}}^{-1/2}(z_2)V_{\text{Gluino}}^{-1/2}(z_3)V_{\text{Gluino}}^{-1/2}(z_4)V_{\text{Gluino}}^{-1/2}(z_5)V_{\text{Gluino}}^{1/2}(z_6)\rangle \nonumber \\
&=\int\frac{d^6z}{SL(2,\mathbb{R})}
\prod_{1\le i<j\le6}z_{i,j}^{-\alpha'c_{i,j}} 
G(z_t,\dot\alpha_t)\,
u^1_{\dot\alpha_1}u^2_{\dot\alpha_2}u^3_{\dot\alpha_3}
u^4_{\dot\alpha_4}u^5_{\dot\alpha_5}u^6_{\dot\alpha_6},\nonumber\\
&=
\int\frac{d^6z}{SL(2,\mathbb{R})\prod_{i=1}^6z_{i,i+1}}
\prod_{1\le i<j\le6}z_{i,j}^{-\alpha'c_{i,j}}\prod_{i=1}^6z_{i,i+1}     
G(z_t,\dot\alpha_t)\,
u^1_{\dot\alpha_1}u^2_{\dot\alpha_2}u^3_{\dot\alpha_3}
u^4_{\dot\alpha_4}u^5_{\dot\alpha_5}u^6_{\dot\alpha_6},
\end{align}
where we identify \(z_{7} \equiv z_1\) in the denominator \(\prod_{i=1}^{6}z_{\,i,i+1}\). Each \(u^k_{\dot\alpha_k}\) is the Weyl spinor wavefunction associated with the \(k\)th gluino (with dotted spinor index \(\dot\alpha_k\)).

The function \( G(z_t, \dot\alpha_t) \) encodes the world-sheet correlator of the six vertex operators, where the label \( \dot\alpha_t \) indicates that this term carries the \( \dot\alpha \) index. One way to express \( G \) is as a sum over intermediate insertions \( i = 1, \dots, 5 \), followed by a sum over permutations \( \pi \in S_5 \)~\cite{Kostelecky:1986ab}. Explicitly,
\begin{equation}
\label{eq:G_definition}
\begin{split}
G(z_t,\dot\alpha_t)
&=\sum_{\pi\in S_5}\mathrm{sign}(\pi)\Biggl\{\sum_{i=1}^5\frac{1}{z_{i,6}}\Bigl(k^{\,i\nu}-\tfrac{1}{8}k_{\mu}^6M_{(i)}^{\,\mu\nu}\Bigl)\Bigl[
-\frac{1}{80}
\frac{A_{\nu}^{\pi,\{\dot\alpha_j\}}}
     {z_{\pi(1),\pi(2)}z_{\pi(2),\pi(3)}z_{\pi(3),\pi(4)}
      z_{\pi(4),\pi(5)}z_{\pi(5),\pi(1)}}
\\[-0.2em]
&\quad+\frac{1}{24}
\frac{B_{\nu}^{\pi,\{\dot\alpha_j\}}z_{\pi(4),6}}
     {z_{\pi(1),\pi(2)}z_{\pi(2),\pi(3)}z_{\pi(3),\pi(4)}
      z_{\pi(4),\pi(5)}z_{\pi(5),6}z_{\pi(4),\pi(1)}}
\Bigr]\Biggr\},
\end{split}
\end{equation}
where
\begin{enumerate}
    \item \(M_{(i)}^{\,\mu\nu}\) acts on the spinor index \(\dot\alpha_i\) as
   \[
   M_{(i)}^{\mu\nu}\bigl(f^{\dot\alpha_6\dot\alpha_5\dot\alpha_4\cdots\dot\alpha_1}\bigr)
   =(\gamma^{\mu\nu})^{\dot\alpha_i}{}_{\dot\beta}\,
   f^{\dot\alpha_6\dot\alpha_5\cdots\dot\alpha_{\,i+1}\dot\beta\dot\alpha_{\,i-1}\cdots\dot\alpha_1},
   \]
   with \(\gamma^{\mu\nu}=\tfrac12[\gamma^\mu,\gamma^\nu]\).
   \item The objects \(A_\nu^{\pi,\{\dot\alpha_t\}}\) and \(B_\nu^{\pi,\{\dot\alpha_t\}}\) are chains of gamma matrices contracted with six spinor indices \(\dot\alpha_1,\ldots,\dot\alpha_6\). In particular,
   \begin{align}
   A^{\pi,\{\dot\alpha_t\}}_{\nu}&:=(\gamma_\nu\gamma^{\rho\lambda})^{\dot\alpha_6,\dot\alpha_{\pi(5)}}(\gamma_\rho)^{\dot\alpha_{\pi(4)},\dot\alpha_{\pi(3)}}(\gamma_\lambda)^{\dot\alpha_{\pi(2)},\dot\alpha_{\pi(1)}},\nonumber\\
   B^{\pi,\{\dot\alpha_t\}}_{\nu}&:=(\gamma_\nu)^{\dot\alpha_6,\dot\alpha_{\pi(5)}}(\gamma^\rho)^{\dot\alpha_{\pi(4)},\dot\alpha_{\pi(3)}}(\gamma_\rho)^{\dot\alpha_{\pi(2)},\dot\alpha_{\pi(1)}}.
   \end{align}
   Here, \(\pi(1),\dots,\pi(5)\) labels a permutation in \(S_5\) acting on \(\{1,2,3,4,5\}\), while index “6” is fixed.
\end{enumerate}

\paragraph{Recombination into Pure \(u\)-Variables.}
To rewrite the two permutation sums in \(G\) as products of \(u\)-variables, we reorganize the spinor indices. Define:
\begin{align}
\bigl(k^{\,i\nu}-\tfrac18k^6_\mu M_{(i)}^{\,\mu\nu}\bigr)\,A_\nu^{\pi,\{\dot\alpha_t\}}u^6_{\dot\alpha_6}
&\equiv C_i^{\pi,\{\dot\alpha_t\}},\nonumber\\
\bigl(k^{\,i\nu}-\tfrac18k^6_\mu M_{(i)}^{\,\mu\nu}\bigr)\,B_\nu^{\pi,\{\dot\alpha_t\}}u^6_{\dot\alpha_6}
&\equiv D_i^{\pi,\{\dot\alpha_t\}}.
\end{align}
Note that while the spinor indices of $C$ and $D$ depends on the permutation, since it is contracted with $u$, the operation is equivalent to holding $C$ and $D$ fixed while performing the ``inverse" permutation:
\begin{align}
\sum_{i=1}^{5}C_i^{\pi,\{\dot\alpha_t\}}\,
u^{5}_{\dot\alpha_{5}}u^{4}_{\dot\alpha_{4}}
u^{3}_{\dot\alpha_{3}}u^{2}_{\dot\alpha_{2}}
u^{1}_{\dot\alpha_{1}}
&=\sum_{i=1}^{5}C_i^{\pi=1,\{\dot\alpha_t\}}\,
u^5_{\dot\alpha_{\pi^{-1}(5)}}u^4_{\dot\alpha_{\pi^{-1}(4)}}u^3_{\dot\alpha_{\pi^{-1}(3)}}u^2_{\dot\alpha_{\pi^{-1}(2)}}u^1_{\dot\alpha_{\pi^{-1}(1)}},\nonumber\\
\sum_{i=1}^{5}D_i^{\,\pi}(\dot\alpha_t)\,
u^{5}_{\dot\alpha_{5}}u^{4}_{\dot\alpha_{4}}
u^{3}_{\dot\alpha_{3}}u^{2}_{\dot\alpha_{2}}
u^{1}_{\dot\alpha_{1}}
&=\sum_{i=1}^{5}D_i^{\pi=1,\{\dot\alpha_t\}}\,
u^5_{\dot\alpha_{\pi^{-1}(5)}}u^4_{\dot\alpha_{\pi^{-1}(4)}}u^3_{\dot\alpha_{\pi^{-1}(3)}}u^2_{\dot\alpha_{\pi^{-1}(2)}}u^1_{\dot\alpha_{\pi^{-1}(1)}}\,.
\end{align}
Then, one can explicitly show:
\begin{equation}
\label{eq:C_property}
\sum_{i=1}^{5}C_i^{\pi=1,\{\dot\alpha_t\}}=0,\,\quad
\sum_{i=1}^{5}D_i^{\pi=1,\{\dot\alpha_t\}}=0\,.
\end{equation}
For convienence, we will suppress the superscript $\{\dot\alpha_t\}$. In the following discussion, we illustrate how to use the identity above to convert the integrand into an expression involving cross-ratios.

We consider a given permutation \( \pi \), and focus on the sum of terms associated with \( A^\pi_\nu \). First, observe that each term associated with \( A^\pi_\nu \) contains the common factor:
\[
\frac{1}{z_{\pi(1),\pi(2)} z_{\pi(2),\pi(3)} z_{\pi(3),\pi(4)} z_{\pi(4),\pi(5)} z_{\pi(5),\pi(1)}},
\]
which carries weight \(-2\) in each of \( z_1, z_2, z_3, z_4, z_5 \), and weight \( 0 \) in \( z_6 \). When combined with the Parke–Taylor factor \( \mathbb{PT} \), the remaining sum must be organized into terms with weight -2 in $z_6$ and weight 0 in other. Indeed  using eq.\eqref{eq:C_property} we can manipulate the sum as follows:
\begin{align}
\sum_{i=1}^5 \frac{C_i^{\,\pi=1}}{z_{i,6}}
&= \sum_{i=1}^4 \left( \frac{C_i^{\,\pi=1}}{z_{i,6}} - \frac{C_i^{\,\pi=1}}{z_{5,6}} \right) \nonumber \\
&= C_1^{\,\pi=1} \frac{z_{1,5}}{z_{1,6}z_{5,6}}
+ C_2^{\,\pi=1} \frac{z_{2,5}}{z_{2,6}z_{5,6}}
+ C_3^{\,\pi=1} \frac{z_{3,5}}{z_{3,6}z_{5,6}}
+ C_4^{\,\pi=1} \frac{z_{4,5}}{z_{4,6}z_{5,6}},
\end{align}
where the coefficients \( C_1^{\,\pi=1}, C_2^{\,\pi=1}, C_3^{\,\pi=1}, C_4^{\,\pi=1} \) are independent of the \( z \)-coordinates. A similar manipulation can be applied to the \( D \)-terms. Thus by reexpanding the amplitude in \( C_1^{\,\pi=1}, C_2^{\,\pi=1}, C_3^{\,\pi=1}, C_4^{\,\pi=1} \), this gives natural $\text{SL(2,\,}\mathbb{R})$ invariant blocks which can be straight forwardly convert into curve-integral representation.\\

To simplify the discusion, let's denote the exponents of \( z_{i,j} \) appearing when rewriting the correlator in terms of blocks associated with \( C_k^{\pi=1} \) and \( D_k^{\pi=1} \) as \( g^{\pi,k}_{i,j} \) and \( h^{\pi,k}_{i,j} \) respectively. Then when rewritten in therms of the $u$-variables, we denote their exponents as \( \beta^{\pi,k}_{m,n} \) and \( \gamma^{\pi,k}_{m,n} \). That is, 
\begin{align}
 &\frac{z_{k,5}}{z_{k,6}z_{5,6}} \cdot \frac{\prod_{i=1}^6 z_{i,i+1}}{z_{\pi(1),\pi(2)} z_{\pi(2),\pi(3)} z_{\pi(3),\pi(4)} z_{\pi(4),\pi(5)} z_{\pi(5),\pi(1)}}
:= \prod_{i<j} z_{i,j}^{g^{\pi,k}_{i,j}} =\prod_{m<n} u_{m,n}^{\beta^{\pi,k}_{m,n}}, 
\nonumber\\[0.5em]
  &\frac{z_{k,5}}{z_{k,6}z_{5,6}} \cdot \frac{\prod_{i=1}^6 z_{i,i+1} \cdot z_{\pi(4),6}}{z_{\pi(1),\pi(2)} z_{\pi(2),\pi(3)} z_{\pi(3),\pi(4)} z_{\pi(4),\pi(5)} z_{\pi(5),6} z_{\pi(4),\pi(1)}}
:= \prod_{i<j} z_{i,j}^{h^{\pi,k}_{i,j}}=\prod_{m<n} u_{m,n}^{\gamma^{\pi,k}_{m,n}}.
\end{align}
Note that each exponent \( g^{\pi,k}_{m,n} \), \( h^{\pi,k}_{m,n} \in \{-1, 0, 1\} \). 
Then using eq.\eqref{eq:u_from_F_T} and \eqref{eq:u_from_z} to convert the \( u \)-variables into $F$-polynomials:
\begin{align}
\prod_{i<j}z_{i,j}^{g^{\pi,k}_{i,j}}=\prod_{m<n} u_{m,n}^{\beta^{\pi,k}_{m,n}}=\prod_{t=1}^3 y_{i_t,j_t}^{\beta^{\pi,k}_{i_t,j_t}}\prod_{i<j}F_{i,j}^{g^{\pi,k}_{ij}},
\nonumber\\
\prod_{i<j}z_{i,j}^{h^{\pi,k}_{ij}}=\prod_{m<n} u_{m,n}^{\gamma^{\pi,k}_{m,n}}=\prod_{t=1}^3 y_{i_t,j_t}^{\gamma^{\pi,k}_{i_t,j_t}}\prod_{i<j}F_{i,j}^{h^{\pi,k}_{ij}}.
\end{align}
Therefore, the six-point gluino amplitude can be rewritten in the following form:
\begin{equation}
\begin{split}
\label{Gluino6}
A^{\text{Gluino}}_6 
&= \int_0^\infty  \prod_{I=1}^{3} \frac{dy_{I}}{y_I} 
    y_I^{\alpha' X_{I}} \prod_{i<j} F_{i,j}^{-\alpha' c_{i,j}} \\
&\quad \times \Bigg[
\sum_{\pi\in S_5} \bigg( 
    \frac{-1}{80} \sum_{k=1}^4 C^{\pi=1}_k \prod_{I=1}^3 y_{I}^{\beta^{\pi,k}_{I}} \prod_{i<j} F_{i,j}^{g^{\pi,k}_{i,j}} 
    + \frac{1}{24} \sum_{k=1}^4 D^{\pi=1}_k \prod_{I=1}^3 y_{I}^{\gamma^{\pi,k}_{I}} \prod_{i<j} F_{i,j}^{h^{\pi,k}_{i,j}}
    \bigg)  \\
&\qquad\qquad 
\times \;
u^5_{\dot\alpha_{\pi^{-1}(5)}} u^4_{\dot\alpha_{\pi^{-1}(4)}}
u^3_{\dot\alpha_{\pi^{-1}(3)}} u^2_{\dot\alpha_{\pi^{-1}(2)}}
u^1_{\dot\alpha_{\pi^{-1}(1)}}
\bigg].
\end{split}
\end{equation}
Here, the first part corresponds to the measure and the Koba–Nielsen factor in eq.\eqref{eq:I6_definition}. This part of rewriting closely resembles the procedure used to derive eq.\eqref{eq: Trphi3y}.

Now, with this new repesentation of gluino amplitude, we can analyze the condition when the amplitude will vanishes. For convenience, we adopt the ray-like triangulation and follow a similar analysis in the super-tachyon case, we must ensure that, for each term in the integrand, all exponents of \( F_{i,j} \) that may depend on a specific \( y_{1,a} \) are non-negative integers. Since the exponent of each \( F_{i,j} \) is of the form \( -\alpha 'c_{i,j} + g^{\pi,k}_{i,j} \) or \( -\alpha'c_{i,j} + h^{\pi,k}_{i,j} \), we require \( \alpha' c_{i,j} \in -\mathbb{N} \) (strictly negative integers) rather than merely non-positive integers.

Concretely, there are three distinct cases under which the integral becomes scaleless:
\begin{align}
&\text{If } \alpha 'c_{1,3},\, \alpha 'c_{1,4},\, \alpha 'c_{1,5} \in -\mathbb{N}, \quad &&\Longrightarrow\quad y_{1,3} \text{ is scaleless}, \nonumber \\
&\text{If } \alpha 'c_{1,4},\, \alpha 'c_{2,4},\, \alpha 'c_{1,5},\, \alpha 'c_{2,5} \in -\mathbb{N}, \quad &&\Longrightarrow\quad y_{1,4} \text{ is scaleless}, \nonumber \\
&\text{If } \alpha 'c_{1,5},\, \alpha 'c_{2,5},\, \alpha 'c_{3,5} \in -\mathbb{N}, \quad &&\Longrightarrow\quad y_{1,5} \text{ is scaleless}.
\end{align}

In each of these cases, the corresponding integration becomes scaleless, and thus every term in the six-point amplitude vanishes.\\

Building on the gluino amplitude expression above, we now derive its curve-integral representation in the field-theory limit (\( \alpha' \to 0 \)). Since $\alpha'$ appears in the exponent, the dominant behaviour of the amplitude in the \( \alpha' \to 0 \) limit is determined by the asymptotic behaviour of the integrand in the limits $y\rightarrow \infty$. To properly capture this limit,  we follow the discussion in~\cite{Arkani-Hamed:2019mrd,Arkani-Hamed:2023lbd}, utilize the notion of tropicalization.
Following the notation in~\cite{Arkani-Hamed:2023lbd}, we denote by \( \operatorname{Trop} F \) the tropicalization of a function \( F \). This process converts rational functions of \( y_i \) into piecewise-linear functions of \( t_i \), via the formal substitutions:  
\begin{itemize}
    \item Variables: \( y_i \mapsto t_i \),
    \item Operations: \( (+,\times,\div) \mapsto (\max,+,-) \).
\end{itemize}
For example,
\[
\operatorname{Trop}\left(\frac{y_1^a y_2^b}{1 + y_1 + y_2}\right)
= at_1 + bt_2 - \max(0, t_1, t_2).
\]
Consider a general string integral of the form:
\begin{equation}
\mathcal{I} := (\alpha')^{n-3} \int_0^{\infty} \prod_{I=1}^{n-3} \frac{dy_{I}}{y_{I}} \,
y_{I}^{\alpha' X_{I}} \prod_{i<j} F_{i,j}^{-\alpha' c_{i,j}}(\mathbf{y}), 
\end{equation}
where we define
\[
R(\mathbf{y}) := \prod_{I=1}^{n-3} y_{I}^{X_{I}} \prod_{i<j} F_{i,j}^{-c_{i,j}}(\mathbf{y}).
\]
Tropicalization allows us to express the curve integral in the \( \alpha' \to 0 \) limit.
Writing 
\[
\operatorname{Trop} R(\mathbf{t}) = \sum_{I=1}^{n-3} X_{I} t_I - \sum_{i<j} c_{i,j} \operatorname{Trop} F_{i,j}(\mathbf{t})\,,
\]
Then we have:
\begin{equation}
\label{scf}
\lim_{\alpha' \to 0} \mathcal{I} =\lim_{\alpha'\to0} \int_{-\infty}^{\infty} (\alpha')^{n-3}\prod_{I=1}^{n-3} dt_I \;
\exp\left( \alpha'\operatorname{Trop} R(\mathbf{t}) \right) 
= \int_{-\infty}^{\infty} \prod_{I=1}^{n-3} dt_I \;
\exp\left( \sum_{I=1}^{n-3} X_{I} t_I - \sum_{i<j} c_{i,j}  f_{i,j}(\mathbf{t}) \right),
\end{equation}
where for simplicity, we denote \( \operatorname{Trop} F_{i,j}(\mathbf{y}) := f_{i,j}(\mathbf{t}) \).


As discussed in~\cite{Arkani-Hamed:2023lbd}, when the integrand is of the form \( \frac{dy}{y} \) and the exponent of each component is proportional to \( \alpha' \), the tropicalization technique  provides a straightforward way to extract the field-theory limit.  However, in the gluino case, these two conditions may fail.
Fortunately, as shown in~\cite{Arkani-Hamed:2023jry}, the \( g \)-vector fan technique allows us to factor out the leading behavior and perform a cone-by-cone expansion to extract the field-theory limit.

To illustrate the idea, we consider the example of the four-point gluino amplitude. In this case, the \( g \)-vector takes values \( \pm 1 \), which defines two cones: \( t \le 0 \) and \( t > 0 \). We perform a change of variable \( y_{1,3} \to e^{t} \). The first term in the integrand is proportional to:
\begin{equation}
\int_0^\infty\frac{dy_{1,3}}{y_{1,3}}y_{1,3}^{\alpha'X_{1,3}}F_{1,3}^{-\alpha'c_{1,3}}=\int_{-\infty}^{\infty} dt\, e^{\alpha'X_{1,3}t} (1 + e^{t})^{-\alpha'c_{1,3}}.
\end{equation}
But we find the above form satisify the form we have discussed, so  we can just apply eq.\eqref{scf} to get the curve-integral form without cone-by-cone expansion:
\begin{equation}
\label{4g1}
 \lim_{\alpha'\rightarrow0}\alpha'\int_0^\infty\frac{dy_{1,3}}{y_{1,3}}y_{1,3}^{\alpha'X_{1,3}}F_{1,3}^{-\alpha'c_{1,3}}=\int^{\infty}_{-\infty}dt e^{X_{1,3}t-c_{1,3}f_{1,3}(t)},
\end{equation}
where $f_{1,3}(t)=\text{max}(0,t)$.
From ~\eqref{4g1}, we find that the total contribution is \( \frac{1}{\alpha'(c_{1,3} - X_{1,3})} +\frac{1}{\alpha'X_{1,3}}\).



The second term in the integrand is proportional to:
\begin{align}
\int_0^\infty\frac{dy_{1,3}}{y_{1,3}}y_{1,3}^{\alpha'X_{1,3}+1}F_{1,3}^{-\alpha'c_{1,3}-1}
&= \int_{-\infty}^{\infty} dt\, e^{\alpha'X_{1,3}t + t} (1 + e^{t})^{-\alpha'c_{1,3} - 1} \nonumber\\
&= \int_{0}^{\infty} dt\, e^{\alpha'X_{1,3}t+t}e^{-\alpha'c_{1,3}t - t} (1+e^{-t} )^{-\alpha'c_{1,3} - 1} \nonumber\\
&\quad + \int_{-\infty}^{0} dt\, e^{\alpha'X_{1,3}t +t}  (1+e^t )^{-\alpha'c_{1,3} - 1}.\nonumber
\end{align}
In the cone \(t> 0\), we expand the remaining factor \( (1 + e^{-t})^{-\alpha'c_{1,3} - 1}\approx 1+(-\alpha'c_{1,3} - 1)e^{-t} \) and find that only the first term 1 might create a contribution $O(\alpha'^{-1})$. This yields an overall contribution of \( \frac{1}{\alpha'(c_{1,3} - X_{1,3})} \). In the other cone \(t\le0\), we also expand the remaining factor in a similar way, \( (1 + e^{t})^{-\alpha'c_{1,3} - 1}\approx 1+(-\alpha'c_{1,3} - 1)e^{t} \). But in this case, each term leads to an integral of order \( \mathcal{O}(1) \), which is subleading in the \( \alpha' \to 0 \) limit.  
Therefore, the total contribution of this part to the field-theory limit is \( \frac{1}{\alpha'(c_{1,3} - X_{1,3})} \).





In summary, the field theory limit of four-point gluino amplitude can be written as the following form:
\begin{equation}
\begin{split}
A^{\text{Gluino,F}}_{4} &=u_\alpha u_\beta u_\lambda u_\delta \bigg[ \frac{-(\gamma^\mu C)_{\alpha\beta}(\gamma_\mu C)_{\lambda\delta}}{2}\left(\frac{1}{(c_{1,3} - X_{1,3})} +\frac{1}{X_{1,3}}\right)\\&\qquad\qquad+\frac{-(\gamma^\mu C)_{\alpha\lambda}(\gamma_\mu C)_{\delta\beta}}{2}\left(\frac{1}{(c_{1,3} - X_{1,3})}\right)\bigg]\\
&=u_\alpha u_\beta u_\lambda u_\delta \bigg[\frac{-(\gamma^\mu C)_{\alpha\beta}(\gamma_\mu C)_{\lambda\delta}}{2}\left(\frac{-1}{u} +\frac{1}{s}\right)+\frac{-(\gamma^\mu C)_{\alpha\lambda}(\gamma_\mu C)_{\delta\beta}}{2}\left(\frac{-1}{u}\right)\bigg].
\end{split}
\end{equation}

In general, this procedure can be applied to the six-point gluino amplitude and extended to higher-point cases. We first analyze the leading order behavior cone-by-cone using tropicalization, after performing the variable change \( y_I \to e^{t_I} \). We leave the analysis of the curve-integral representation and its field theory limit for general $2n$-point for the future.

\subsection{From Gluino to Yang-Mills amplitude}
Not surprisingly, one can similarly obtain  super Yang-Mills amplitude via scaffolding gluino amplitudes. We proceed by taking two gluino vertex operators  and identify:
\[
p_{1} + p_{2} = k, \quad \frac{1}{\sqrt{2}}u_\alpha (\gamma^\mu C)u_\beta=\epsilon^{\mu}, \quad \epsilon\cdot k=0,
\]
where $c_{1,2}=0$. We consider two combinations: \((-\tfrac{1}{2}, -\tfrac{1}{2})\), and \((-\tfrac{1}{2}, \tfrac{1}{2})\), which yield Yang–Mills vertex operators in picture \(-1\), and \(0\), respectively.

\paragraph{Case 1: {Gluino \((-\tfrac{1}{2}, -\tfrac{1}{2}) \rightarrow V_{\rm{SYM}}^{-1}\)}} 

\begin{equation}
\begin{split}
    &V_{\text{Gluino}}^{-1/2}(u_{\alpha},p_2,z_2)V_{\text{Gluino}}^{-1/2}(u_\beta ,p_1,z_1) = u^\alpha \Theta_\alpha e^{-\phi/2} e^{ip_2\cdot X}(z_2)\,\,u^\beta \Theta_\beta e^{-\phi/2} e^{ip_1\cdot X}(z_1)   \\
    & \sim u_\alpha u_\beta\left( \frac{1}{z_{1,2}^{1/4}}\right)\left(\frac{(\gamma^\mu C)_{\alpha\beta} \psi_\mu}{\sqrt{2}z_{1,2}^{3/4}}\right) e^{-\phi} e^{ik\cdot X}(z_1)
\end{split}
\end{equation}
Taking the residue yields a vertex operator of the form:
\begin{equation}
    V_{\rm{SYM}}^{-1}=(\epsilon \cdot \psi) e^{-\phi}e^{ik\cdot X}.
\end{equation}

\paragraph{Case 2: {Gluino \((-\tfrac{1}{2}, \tfrac{1}{2}) \rightarrow V_{\rm{SYM}}^{0}\)}}

Instead of using the explicit form of \( V_{\text{Gluino}}^{1/2} \) to evaluate the OPE, it is much easier to use the picture-changing operator:

\begin{equation}
\begin{split}
    &V_{\text{Gluino}}^{-1/2}(u_{\alpha},p_2,z_2)V_{\text{Gluino}}^{1/2}(u_\beta ,p_1,z_1) = u^\alpha \Theta_\alpha e^{-\phi/2} e^{ip_2\cdot X}(z_2)\,\,\left(-2\left[ Q_\text{BRST},\, \xi(z)V_{\text{Gluino}}^{-1/2}(u,p,z) \right] \right)(z_1)   \\
    & = u^\alpha \Theta_\alpha e^{-\phi/2} e^{ip_2\cdot X}(z_2)\,\,\Bigg(-2\oint \frac{dw}{2\pi i}e^\phi\eta \,i\partial X\cdot\psi (w)\Bigg)\, \xi(z_1)u^\beta \Theta_\beta e^{-\phi/2} e^{ip_1\cdot X}(z_1) \\
    &\sim u_\alpha u_\beta\left(\frac{(\gamma^\mu C)_{\alpha\beta} }{\sqrt{2}z_{1,2}}\right)  (i\partial X_\mu+2\alpha'(k\cdot\psi)\psi_\mu)e^{ik\cdot X}(z_1)
\end{split}
\end{equation}
Taking the residue yields a vertex operator of the form:
\begin{equation}
    \Rightarrow V_{\rm{SYM}}^{0}=((\epsilon\cdot i\partial X)+2\alpha'(k\cdot\psi)(\epsilon\cdot\psi)) e^{ik\cdot X}.
\end{equation}

The scaffolding of the \( 2n \)-point gluino amplitude thus follows:

\begin{equation}
A^{\text{Gluino}}_{2n} 
= \int \frac{\prod_{a=1}^{2n} dz_{2a}}{\mathrm{SL}(2,\mathbb{R})}\,\Bigl\langle\prod_{a=1}^{2n} V_{\text{Gluino}}^{q_a}(z_a)\Bigr\rangle,
\end{equation}
\begin{equation}
\rightarrow A^{\text{SYM}}_{n} 
= \int \frac{\prod_{i=1}^n dz_{i}}{\mathrm{SL}(2,\mathbb{R})}\,\Bigl\langle\prod_{i=1}^{n} V_\text{{SYM}}^{q_i}(z_i)\Bigr\rangle,
\end{equation}
by considering the factorization poles given by \( X_{2i-1,2i} := (p_{2i-1} + p_{2i})^2 = k_i^2 = 0 \), and the superghost charge conservation should also be satisfied before and after the scaffolding procedure.

\section{Conclusions}

In this paper, we've considered curve intergral representation for general string amplitude, which can be iteratively derived from the seed amplitudes: tachyon, super-tachyon or gluino amplitudes for bosonic, NS-sector and R-sector string theory. Due to the simple relation between $F$-polynomials of the pre-scaffold and descendents, the zeros of string amplitudes with any level can be straight forwardly deduced. Importantly, $n$-point higher level amplitudes enjoys an enhanced set of zeros: the ones that are shared with the $n$-point tachyon amplitude and those that are inherited from its $(2^N)n$ parent tachyon amplitude. 

We have also extended our analysis to open superstring amplitudes, both for the NS and R-sectors. Note that from either sector one can straightforwardly scaffold super Yang-Mills level-1 amplitude, which we demonstrate on the vertex algebra level. The super-tachyon amplitude, which is the parent for the NS-sector, is closely related to the tachyon amplitude simply by replacing $z_{i,j}$ with its "superspace" counter-part:
\begin{equation}
\int \frac{d^{n} z_i}{\text{SL(2,\,}\mathbb{R})}\prod_{i<j}z^{-\alpha' c_{i, j}} \rightarrow \int \frac{d^{n} z_id^{2n-2}\theta}{\text{SL(2,\,}\mathbb{R})}\frac{1}{z_{a,b}}\prod_{i<j}|z_{i,j}-\theta_i\theta_j|^{-\alpha' c_{i, j}}\,.
\end{equation}
It will be interesting if there is a notion of super-$u$ variables, where one defines:
\begin{equation}
\tilde{u}_{i,j}\equiv \frac{|z_{i{-}1,j}-\theta_{i{-}1}\theta_{j}||z_{i,j{-}1}-\theta_{i}\theta_{j{-}1}|}{|z_{i,j}-\theta_i\theta_j||z_{i{-1},j{-}1}-\theta_{i{-}1}\theta_{j{-}1}|}.
\end{equation}
These $\tilde{u}$-variables will then satisfy: 
\begin{equation}\label{eq: ueq}
\tilde{u}_{i,j}+\prod_{(k,l)\cap (i,j)}\tilde{u}_{k,l}=1{+}{\rm fermion\;terms}\,.
\end{equation}
It will be interesting to explore this further. Note that both for the NS- and R-sector the seed amplitude on-top of the scaffolding tower, the super-tachyon and gluino amplitude, is even-multiplicity in nature. This suggests the supersymmetrization of surfaceology would entail fat-graphs built on quartic diagrams.  

Given the curve-integral representation of the gluino amplitude, it is straight forward to take the field theory limit to reproduce the colored fermion amplitudes in the curve-representation. It will be desirable to obtain a closed form for general $2n$-point, purhaps bowrrowing insights from the Yukawa fermion~\cite{De:2024wsy}. Finally, in the field theory limit, the presence of zeros can be related to enhanced ultraviolet scaling under BCFW shifts~\cite{Rodina:2024yfc}. It will be interesting to study whether there is similar connection with stringy BCFW shifts~\cite{Cheung:2010vn}.

\textbf{Acknowledgments}
We would like to thank Laurentiu Rodina, Song He, Qu Cao and Jing Dong for useful comments. 
Yu-Chi Chang, Hsing-Yen Chen and Y.-t. Huang is supported by the Taiwan
Ministry of Science and Technology Grant No. 112-2628-M-002-003-MY3. 

\appendix
\section{Scaffolding Super Yang-Mills on correlators}\label{sec: App}

We have shown in sec.\ref{yangmillsvertexscaffold} that the super Yang-Mills vertex operators can be obtained by taking the residue for the OPE of two super-tachyon vertex operators. Here we explicitly show that this is equivalent to doing the scaffolding on the correlator level from the super-tachyon amplitudes to get the resulting super Yang-Mills amplitudes.

Begin with the super-tachyon amplitude:
\begin{equation}
A^{\text{ST}}_{2n} 
= \int \frac{\prod_{a=1}^{2n} dz_{2a}}{\mathrm{SL}(2,\mathbb{R})}\,\Bigl\langle\prod_{a=1}^{2n} V_\text{{ST}}^{q_a}(z_a)\Bigr\rangle,
\end{equation}
where \(V_\text{{ST}}^{-1}(z) = e^{-\phi} e^{ip\cdot X} (z),\, V_\text{{ST}}^{0}(z) = p\cdot\psi e^{ip\cdot X} (z)\,,\) $p^2 = \frac{1}{2\alpha'}$, and $\sum_aq_a=-2$, i.e. we have 2 picture $-1$ and $n{-}2$ picture $0$. There are two possible ways to take the residues: either the two picture $-1$ vertex operators merge together, or each picture $-1$ vertex operator takes the residue with a different picture 0 vertex operator, assuming that we take the residues at $z_{2a-1,2a}$. Since the amplitude is independent of the choice of picture fixing, the result of these two choices are guaranteed to be equivalent. After merging, we identify the momenta of the operators as $p_{2a-1} + p_{2a} = k_i$, $p_{2a}=\epsilon$, and $c_{2a-1,2a}=\tfrac{1}{\alpha'}$.

Let's consider the first case, namely, where we scaffold two picture $-1$ vertex operators and the remaining pitcture $0$. This corresponds to taking the residue on $z_{1,2}$ and $z_{3,4}$ for the following correlator:
\begin{equation}
    \Bigl\langle e^{-\phi}e^{ip_1\cdot X}(z_1)e^{-\phi}e^{ip_2\cdot X}(z_2)p_3\cdot \psi e^{ip_3\cdot X}(z_3)p_4\cdot \psi e^{ip_4\cdot X}(z_4)\prod_{a=4}^{2n} V_{\text{ST}}^{0}(z_a)\Bigr\rangle.
\end{equation}
The analysis will be similar to sec.\ref{scaffoldcorrelator},  \( z_{1,2} \) is given by:
\begin{equation}
\left( \frac{1}{z_{1,2}^2} e^{-\phi}e^{ip_1\cdot X}(z_1)e^{-\phi}e^{ip_2\cdot X}(z_2)\right),
\end{equation}
where the exponential factors can only contract with other vertices.  The dependence on \( z_{3,4} \) is 
\begin{equation}
\left( \frac{p_3 \cdot p_4}{z_{3,4}^2}e^{ip_3\cdot X}(z_3)e^{ip_4\cdot X}(z_4) ,\quad \frac{(p_3\cdot \psi)(p_4\cdot\psi)}{z_{3,4}}e^{ip_3\cdot X}(z_3)e^{ip_4\cdot X}(z_4)\right).
\end{equation}
The analysis for the remaining \( z_{2a-1,2a} \) is the same as that for \( z_{3,4} \), so we use \( z_{3,4} \) as an example.
Then, take the residue and set $z_1=z_2$, $z_3=z_4$, $z_{2a-1}=z_{2a}$, and relabel the indices, we obtain:
\begin{equation}
\begin{split}
    &\Bigl\langle (\epsilon_1 \cdot  i\partial X -\partial\phi) e^{-2\phi}e^{ik_1\cdot X}(z_1)\\&\qquad\qquad ((\epsilon_2\cdot i\partial X)+2\alpha'(k_2\cdot\psi)(\epsilon_2\cdot\psi)) e^{ik_2\cdot X}(z_2)\prod_{i=3}^{n} V_{\text{SYM}}^{0}(z_i)\Bigl\rangle,
\end{split}
\end{equation}
which corresponds to the super Yang-Mills correlator with one picture -2 vertex and all others in picture 0.

For the second case, where each picture -1 vertex operator takes the residue with a different picture 0 vertex operator, the correlator is given by:
\begin{equation}
    \Bigl\langle e^{-\phi}e^{ip_1\cdot X}(z_1)p_2\cdot \psi e^{ip_2\cdot X}(z_2)e^{-\phi}e^{ip_3\cdot X}(z_3)p_4\cdot \psi e^{ip_4\cdot X}(z_4)\prod_{a=4}^{2n} V_\text{ST}^{0}(z_a)\Bigr\rangle.
\end{equation}
Again, we check the dependence of $z_{1,2}$, which is
\begin{equation}
\left(\frac{p_1\cdot p_2}{z_{12}}e^{-\phi}e^{ip_1\cdot X}(z_1)e^{ip_2\cdot X}(z_2)\right),
\end{equation}
After taking all the residues, we obtain:
\begin{equation}
    \Bigl\langle (\epsilon \cdot \psi) e^{-\phi}e^{ik_1\cdot X}(z_1)(\epsilon \cdot \psi) e^{-\phi}e^{ik_2\cdot X}(z_2)\prod_{i=3}^{n} V_{\text{SYM}}^{0}(z_i)\Bigl\rangle,
\end{equation}
which corresponds to the super Yang-Mills correlator with two picture -1 vertex and all others in picture 0.


\bibliographystyle{JHEP}
\bibliography{biblio}

\end{document}